\documentclass[%
 reprint,
 prd,
 amsmath,amssymb,
 aps,
 lengthcheck,
floatfix,
]{revtex4}

\usepackage{graphicx}% Include figure files
\usepackage{dcolumn}% Align table columns on decimal point
\usepackage{bm}% bold math
\usepackage{hyperref}% add hypertext capabilities
\usepackage[utf8]{inputenc}
\usepackage{multirow}
\usepackage{soul}
\usepackage{float}
\usepackage{graphicx}
\usepackage{times}
\usepackage[normalem]{ulem}
\usepackage{color}
\usepackage{cellspace}
\usepackage[usenames,dvipsnames]{xcolor}

\usepackage{lipsum}% http://ctan.org/pkg/lipsum

\newcommand{\ms}{\scriptscriptstyle}
\newcommand{\be}{\begin{equation}}
\newcommand{\ee}{\end{equation}}
\newcommand{\bea}{\begin{eqnarray}}
\newcommand{\eea}{\end{eqnarray}}
\newcommand {\non}{\nonumber\\}
\newcommand {\nonu}{\nonumber}

\newcommand{\comment}[1]{}
\renewcommand\sout{\bgroup \color{red} \ULdepth=-.5ex \ULset}
\def\simge{\mathrel{\rlap{\raise 0.511ex
     \hbox{$>$}}{\lower 0.511ex \hbox{$\sim$}}}}
\def\simle{\mathrel{\rlap{\raise 0.511ex
      \hbox{$<$}}{\lower 0.511ex \hbox{$\sim$}}}}

\begin{document}

% \setcounter{page}{1}
% \vspace*{0.3 true in}

\title{Nearly model-independent constraints on  dense matter equation of state in a Bayesian approach}

\author{\href{https://orcid.org/0000-0003-0103-5590}N. K. Patra$^{1}$\includegraphics[scale=0.06]{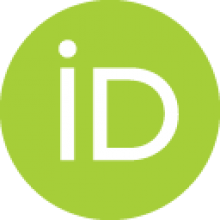}}

\author{\href{https://orcid.org/0000-0003-3308-2615}Sk Md Adil Imam$^{2,3}$\includegraphics[scale=0.06]{Orcid-ID.png}}

\author{\href{https://orcid.org/0000-0001-5032-9435}B. K. Agrawal$^{2,3}$\includegraphics[scale=0.06]{Orcid-ID.png}}
\email{bijay.agrawal@saha.ac.in}

\author{\href{https://orcid.org/0000-0003-1274-5846}Arunava Mukherjee $^{2,3}$\includegraphics[scale=0.06]{Orcid-ID.png}}

\author{\href{https://orcid.org/0000-0003-2633-5821}Tuhin Malik$^4$\includegraphics[scale=0.06]{Orcid-ID.png}}

\affiliation{$^1$Department of Physics, BITS-Pilani, K. K. Birla Goa
Campus, Goa 403726, India}
\affiliation{$^2$Saha Institute of Nuclear Physics, 1/AF 
Bidhannagar, Kolkata 700064, India.}  
\affiliation{$^3$Homi Bhabha National Institute, Anushakti Nagar, Mumbai 400094, India.}  
\affiliation{$^4$CFisUC, Department of Physics, University of Coimbra,
3004-516 Coimbra, Portugal}

\date{\today}

\begin{abstract} 
We apply Bayesian approach to construct a large  number of minimally constrained equations of state (EOSs) and study their correlations with a few selected properties of a neutron star (NS). Our set of minimal constraints includes a few basic properties of saturated nuclear matter and  low-density pure neutron 
matter EOS which is obtained from a precise next-to-next-to-next-to-leading-order (N$^{3}$LO) calculation in
chiral effective field theory. The tidal deformability and radius  of NS with mass $1-2 M_\odot$ are found to be strongly correlated with the pressure of $\beta$-equilibrated matter at densities higher than the saturation density ($\rho_0 = 0.16$ fm$^{-3}$) in a nearly model-independent manner. These correlations are employed to parametrize the pressure for $\beta$-equilibrated matter, around 2$\rho_0$, as a function of neutron star mass and the corresponding tidal deformability. The maximum mass of neutron star is also found to be  strongly correlated with  the pressure of $\beta$-equilibrated matter at densities $\sim 4.5\rho_0$.

\end{abstract}

%\pacs{21.30.Fe, 21.65.Cd, 21.65.Mn, 21.65.Ef}
%\keyword{Neutron Star, Equation of State, $\beta$-Equilibrium Matter, Symmetric Nuclear Matter,  Bayesian Parameter Estimation}

\maketitle
\section{Introduction}
Gravitational-wave astronomy promises unprecedented constraints on the Equation of State of neutron star matter through the detailed properties of gravitational-waveform observed during the merging of binary neutron stars (BNS). In addition, X-ray observations from the Neutron star Interior Composition Explorer (NICER) instruments have also provided a compelling constraint on the equation of state independently. The tidal deformability parameters inferred from these gravitational-wave events encode information about the EOS. For the first time,a BNS event (GW170817) was observed by the LIGO-Virgo detector from a low mass compact binary neutron star merger with a total mass of the system $2.74_{-0.01}^{+0.04}M_\odot$ \cite{Abbott18a, Abbott2019}. Another gravitational-wave event likely originating from the coalescence of BNSs, GW190425, was observed \cite{Abbott2020} subsequently. These two events have already triggered many theoretical investigations to constrain the EOS of neutron star matter \cite{GW170817,Malik2018,De18,Fattoyev2018a,Landry2019,Piekarewicz2019,Malik:2019whk,Biswas2020,Abbott2020,Thi2021}. The upcoming runs of LIGO-Virgo-KAGRA and the future detectors, e.g., Einstein Telescope (ET) and Cosmic Explorer (CE), are expected to observe many more BNS signals emitted from coalescing neutron stars. The mass and radius of NS, observed either in isolation or in binaries, by  NICER \cite{Watts2016, Miller2019, Riley2019} have offered complementary constraints on the EOS. A sufficiently large number of such observations over a wide range of NS masses may be employed to constrain several key quantities associated with the EOS of $\beta$-equilibrated matter which are not readily accessible in the terrestrial laboratory. The behavior of the EOS at supra-saturation densities are generally studied using the  observed maximum neutron star mass, together with radius and tidal deformability corresponding to the neutron star with canonical mass $ 1.4M_\odot$  \cite{Zhang2018, Cai2021,Gil:2021ols}. Recently, in Refs. \cite{Lattimer:2000nx,Maselli:2013mva,Abbot2018,Lim:2018bkq,Tsang:2019vxn,Tsang:2020lmb},  efforts are made to constrain the EOS of $\beta$-equilibrated matter which is relevant to the studies of NS properties. The values of tidal deformability of NS with mass $1-2 M_\odot$ are found to be strongly correlated with the EOS at twice the saturation density.

Statistical tools are quite helpful in providing a quantitative interpretation of NS observables. A Bayesian approach is often applied to analyze gravitational-wave signals, which involves nearly 15 parameters for binary compact object mergers, to infer their source properties \cite{Ashton2019}. It has been also extended to investigate the properties of short gamma-ray bursts  \cite{Biscoveanu2020a}, neutron stars \cite{Coughlin2019,HernandezVivanco2019,Biscoveanu2019a}, the formation history of binary compact objects \cite{Lower2018,Romero-Shaw2019,Ramos-Buades2020,Romero-Shaw2020a,Zevin2020} and to test general relativity \cite{Keitel_2019,Ashton2020,Payne2019,Zhao2019}. Of late,  Bayesian approach has become a useful statistical tool for parameter estimation in the field of nuclear physics  and nuclear-astrophysics \cite{Wesolowski2016}. It allows one to
obtain joint posterior distributions of the model parameters and the correlations among them for a given set of data.  Various constraints on the parameters known $\it {a }$ $\it{priori}$ are incorporated through their prior distributions. The Bayesian techniques
have also been employed to constrain symmetry energy \cite{Somasundaram2021}, masses and
radii of NSs \cite{Drischler2021} using the bounds on the EOS obtained from chiral effective
field theory. Bayesian techniques have been extensively applied to constrain the EOS for symmetric nuclear matter, $\beta$-equilibrated matter (BEM) and density dependence of symmetry energy coefficient using various  finite nuclei and NS properties \cite{Tews2017,Carreau2019,Thi2021a,Riley2018,Raaijmakers:2019dks,Jiang:2019rcw,Guven2020,Biswas2021a,Landry2020a,Huth2021,Biswas2021b,Imam:2021dbe}.

We use Bayesian approach to construct large sets of EOSs which correspond to the Taylor and $\frac{n}{3}$ expansions\cite{Imam:2021dbe}. The expansion coefficients in the former case are the individual nuclear matter parameters (NMPs), whereas in the latter case they are their linear combinations. The EOSs are consistent with a set of minimal constraints that includes a few low-order nuclear matter parameters at the saturation density and  EOS for the pure neutron matter (PNM) at  low densities obtained from a precise next-to-next-to-next-to-leading-order (N$^{3}$LO) calculation in chiral effective field theory.  The  marginalized posterior distributions of NMPs and the various NS properties obtained from set of minimal constraints are found to be within  reasonable bounds. The correlations of various NS  properties, such as tidal deformability, radius and maximum mass, with  key EOS
parameters are studied. These correlations are investigated for a wide range of NS masses and densities for the EOS.

The paper is organized as follows. The Taylor and $\frac{n}{3}$ expansions for the EOS of neutron star matter and the Bayesian approach are briefly outlined in Sec. \ref{methe}. The results for the posterior distributions of nuclear matter parameters and associated NS properties together with their correlations with some key quantities associated with EOS are presented in Sec. \ref{results}. The main outcomes of the present investigation are summarized in Sec. \ref{summary}. 

%%%%%%%%%%%%%%%%%%%%%%%%%%%%%%%%%%%%%%%%%%
\section{Methodology} 
\label{methe}
The   energy per nucleon
 for neutron star matter $E(\rho, \delta)$ at a given  total
nucleon density $\rho$ and asymmetry $\delta$ can be decomposed into the
energy per nucleon for the symmetric nuclear matter, $E(\rho,0)$
and the  density-dependent symmetry energy, $E_{\rm sym}(\rho)$ in the  parabolic approximation as,

\bea
E(\rho, \delta) &=&  E(\rho,0)+E_{\rm sym}(\rho)\delta^2
+..., \label{eq:EOS}
\eea
where,  $\delta  = \left(\frac{\rho_n -\rho_p}{\rho}\right )$ with
$\rho_n$ and  $\rho_p$ being the neutron and  proton densities,
respectively. The value of $\delta$ at a given $\rho$ is determined by
the condition of $\beta$-equilibrium and the charge neutrality. Once
$\delta$ is known, the fraction of neutrons, protons, electrons, muons can be
easily evaluated.  In the following, we expand $E(\rho,0)$ and
$E_{\rm sym}(\rho)$ appearing in Eq. (\ref{eq:EOS})  using Taylor and $\frac{n}{3}$
expansions.  The coefficients of expansion in case of the Taylor
correspond to the individual nuclear matter parameters. In the latter
case, they are expressed as linear combinations of the nuclear matter
parameters. 

\subsection{{Taylor's expansion}}

The $E(\rho,0)$ and $E_{\rm sym}(\rho)$ can be
expanded around the saturation density $\rho_0$ as  \cite{Chen:2005ti,Chen:2009wv,Newton:2014iha,Margueron:2017eqc,Margueron:2018eob},

\bea
E(\rho, 0)&=&	\sum_{n} \frac{a_n}{n!}\left(\frac{\rho-\rho_0}{3\rho_0}\right )^n, \label{eq:SNM_T} \\ 
E_{\rm sym}(\rho) &=&	\sum_{n} \frac{b_n}{n!}\left(\frac{\rho-\rho_0}{3\rho_0}\right )^n, \label{eq:sym_T}  
\eea
so that, 
\bea
E(\rho, \delta)&=&	\sum_{n} \frac{1}{n!}(a_n + b_n\delta^2)\left(\frac{\rho-\rho_0}{3\rho_0}\right )^n ,\label{eq:ebeta_T} 
\eea
where the coefficients $a_n$ and $b_n$ are the nuclear matter
parameters. We truncate the sum in Eqs. (\ref{eq:SNM_T}) and
(\ref{eq:sym_T}) at fourth order, i.e., $n = 0$ - $4$.  Therefore, the
coefficients $a_n$ and $b_n$ correspond,

\bea
a_n & \equiv & \varepsilon_0, 0, K_0, Q_0, Z_0\label{eq:anm1}, \\
b_n  &\equiv & J_0, L_0, K_{\rm sym,0},Q_{\rm sym,0}, Z_{\rm sym,0}. \label{eq:bnm1}
\eea
In Eqs. (\ref{eq:anm1}) and (\ref{eq:bnm1}), $\varepsilon_0 $ is the binding energy per nucleon, $K_0$ {the} incompressibility coefficient, $J_0$ {the} symmetry energy coefficient, its slope parameter $L_0$, $K_{\rm sym,0}$ {the} symmetry energy curvature parameter, $Q_0 (Q_{\rm sym,0})$ and $Z_0 (Z_{\rm sym,0})$ are related to third- and fourth-order density derivatives of  $E (\rho,0) $ [ $E_{\rm sym}(\rho)$], respectively. The subscript zero indicates that all the nuclear matter parameters are calculated at the saturation density.

It may be noticed from Eq. (\ref{eq:ebeta_T}) that the coefficients $a_n$ and $b_n$ may display some correlations among themselves provided the asymmetry parameter depends weakly on the density.  Furthermore, Eq. (\ref{eq:ebeta_T}) may converge slowly at high densities, i.e., $\rho \gg 4\rho_0$. This situation is encountered for the heavier neutron stars. Neutron stars with a mass around $2M_\odot$, typically have central densities $\sim 4-6\rho_0$.

\subsection{$ {\frac{n}{3}}$ expansion}
An alternative expansion of $E(\rho,\delta)$ can be obtained by expanding $E(\rho,0)$ and
$E_{\rm sym}(\rho)$ as \cite{Lattimer2015,Gil2017},

\bea
E(\rho,0) &=& \sum_{n=2}^6 (a'_{n-2}) \left(\frac{\rho }{\rho_0}\right )^{\frac{n}{3}},\label{eq:SNM_n3}\\
E_{\rm sym}(\rho) &=& \sum_{n=2}^6 (b'_{n-2}) \left(\frac{\rho }{\rho_0}\right )^{\frac{n}{3}},\label{eq:sym_n3}\\
E(\rho, \delta) &=& \sum_{n=2}^6 (a'_{n-2} +b'_{n-2} \delta^2 ) \left(\frac{\rho }{\rho_0}\right )^{\frac{n}{3}}.\label{eq:ebeta_n3}
\eea
 We refer {this} as {the} ${\frac{n}{3}}$ expansion. It is now evident from Eqs. (\ref{eq:SNM_n3}) and (\ref{eq:sym_n3}) that the coefficients of expansion are no longer the individual nuclear matter parameters unlike in the case of Taylor's expansion.
The values of the nuclear matter parameters can be expressed in terms of the expansion coefficients $a'$ and $b'$ as, respectively,
\bea
\left (
\begin{matrix}
\varepsilon_0\\
0\\
K_0 \\
Q_0\\
Z_0\\
\end{matrix}
\right ) &=& 
\left (
\begin{matrix}
1 &1 &1 & 1&1\\
2 &3 &4 & 5&6\\
-2  &  0 & 4 & 10&18\\
8  &  0 & -8 & -10 & 0\\
-56 & 0 & 40 & 40 & 0\\
\end{matrix}
\right )
\left (
\begin{matrix}
a'_0\\
a'_1\\
a'_2\\
a'_3\\
a'_4\\
\end{matrix}
\right ),\label{mat_a}\\
\left (
\begin{matrix}
J_0\\
L_0 \\
K_{\rm sym,0}\\
Q_{\rm sym,0}\\
Z_{\rm sym,0}\\
\end{matrix}
\right ) &=& 
\left (
\begin{matrix}
1 &1 &1 & 1&1\\
2  &  3 & 4 & 5&6\\
-2  &  0 & 4 & 10&18\\
8 & 0 & -8 & -10 & 0\\
-56 & 0 & 40 & 40 & 0\\ 

\end{matrix}
\right )
\left (
\begin{matrix}
b'_0\\
b'_1\\
b'_2\\
b'_3\\
b'_4\\
\end{matrix}
\right ).\label{mat_b}
\eea
The relations between the expansion coefficients and the nuclear matter parameters are governed by the nature of functional form for $E(\rho,0)$ and $E_{\rm sym}(\rho)$. The off-diagonal elements in the above matrices would vanish for the Taylor expansion of  $E(\rho,0)$ and $E_{\rm sym}(\rho)$ as given by Eqs. (\ref{eq:SNM_T}) and (\ref{eq:sym_T}), respectively. Therefore, each of the expansion coefficients is simply the individual nuclear matter parameter given by Eqs.  (\ref{eq:anm1}) and (\ref{eq:bnm1}). Inverting the matrices in Eqs.  (\ref{mat_a}) and (\ref{mat_b}) we have
\bea
a'_0&=& \frac{1}{24}(360 \varepsilon_0 + 20K_0   + Z_0),\nonu\\
a'_1&=& \frac{1}{24}(-960 \varepsilon_0 - 56K_0  - 4Q_0 - 4Z_0),\nonu\\
a'_2&=& \frac{1}{24}( 1080\varepsilon_0 + 60K_0 + 12Q_0 + 6Z_0),\nonu\\
a'_3&=& \frac{1}{24}(-576\varepsilon_0 - 32K_0 - 12Q_0 - 4Z_0),\nonu\\
a'_4&=& \frac{1}{24}( 120\varepsilon_0 + 8K_0 + 4Q_0  + Z_0),\label{eq:anm2}\\
b'_0&=& \frac{1}{24}(360J_0 -120L_0  + 20K_{\rm {sym,0}} + Z_{\rm {sym,0}}),\nonu\\
b'_1 &=& \frac{1}{24}(-960J_0 + 328L_0 - 56K_{\rm {sym,0}} - 4Q_{\rm {sym,0}}\nonumber  \\
&&    -4Z_{\rm {sym,0}}),\non
b'_2 &=& \frac{1}{24}(1080J_0 - 360L_0 + 60K_{\rm {sym,0}} + 12Q_{\rm {sym,0}} \nonumber \\
&&   + 6Z_{\rm {sym,0}}),  \non
b'_3 &=& \frac{1}{24}(-576J_0 + 192L_0 - 32K_{\rm {sym,0}} - 12Q_{\rm {sym,0}}  \nonumber \\
&&   -4Z_{\rm {sym,0}}),\non
b'_4 &=& \frac{1}{24}(120J_0 - 40L_0 + 8K_{\rm {sym,0}}  + 4Q_{\rm {sym,0}}  \nonumber \\
&&   + Z_{\rm {sym,0}}).\label{eq:bnm2}
\eea
Each of the coefficients $a'$ and $b'$  are the linear combinations of
nuclear matter parameters in such a way that the lower-order parameters
may contribute dominantly at low densities. The effects of higher-order parameters become prominent with the increase in density.

\subsection{ Bayesian estimation of nuclear matter parameters}\label{BA}

A Bayesian approach enables one to carry out detailed statistical analysis
of the parameters of a model for a given set of fit data. It yields
joint posterior distributions of model parameters which can be used not only to study the distributions of given parameters but also to examine correlations among model parameters. One
can also incorporate prior knowledge of the model parameters and various
constraints on them through the prior distributions. This approach  is
mainly based on the Bayes theorem which states  that \cite{Gelman2013},

\begin{equation}
P(\bm{\theta} |D ) =\frac{{\mathcal L } (D|\bm{\theta}) P(\bm {\theta })}{\mathcal Z},\label{eq:bt}
\end{equation}
where $\bm{\theta}$  and $D$ denote the set of model parameters and
the fit data, respectively. The $P(\bm{\theta} |D )$ is the joint posterior distribution
of the parameters, $\mathcal L (D|\bm{\theta})$ is the likelihood function, $
P(\bm {\theta })$ is the prior for the  model parameters and $\mathcal Z$
is the evidence. The posterior distribution  of a given parameter can be
obtained by marginalizing $P(\bm{\theta} |D )$ over remaining parameters.
The marginalized posterior distribution for a  parameter $\theta_i$
can be obtained as,
\begin{equation}
 P (\theta_i |D) = \int P(\bm {\theta} |D) \prod_{k\not= i }d\theta_k. \label{eq:mpd}
\end{equation}
We use Gaussian likelihood function defined as, 
\bea
{\mathcal L} (D|\bm{\theta})&=&\prod_{j} 
\frac{1}{\sqrt{2\pi\sigma_{j}^2}}e^{-\frac{1}{2}\left(\frac{d_{j}-m_{j}(\bm{\theta)}}{\sigma_{j}}\right)^2}. 
\label{eq:likelihood}  
\eea
Here the index $ j$ runs over all the data, $d_j$ and $m_j$ are the data
and corresponding model values, respectively.  The $\sigma_j$ are the adopted
uncertainties.  The evidence $\mathcal Z$ in Eq. (\ref{eq:bt}) is obtained
by complete marginalization of the likelihood function. It is relevant when
employed to compare different models. However in the present work $\mathcal
Z$ is not very relevant.  To populate the posterior
distribution of Eq. (\ref{eq:bt}), we implement a nested sampling algorithm
by invoking the Pymultinest nested sampling  \cite{Buchner2014} in the
Bayesian Inference Library  \cite{Ashton2019}.

\section{Results and Discussions} \label{results}

We obtained the EOSs for $\beta$-equilibrated matter (BEM)  using Taylor and $\frac{n}{3}$ expansions as discussed in previous section Eqs. (\ref{eq:ebeta_T}) and (\ref{eq:ebeta_n3}). The  coefficients  of  the Taylor expansion  are the individual nuclear matter parameters, whereas, they correspond to linear combinations of nuclear matter parameters for the $\frac{n}{3}$ expansion.    
We have constructed marginalized posterior distributions for the nuclear matter parameters  by applying a Bayesian approach to both the  expansions considered. The  nuclear matter parameters  or the corresponding EOSs are  consistent with  a set of
minimal constraints that includes basic properties of saturated nuclear
matter  and low-density ($\rho=0.08-0.16$fm$^{-3}$) EOS for the pure neutron matter  from  (N$^{3}$LO) calculation in chiral effective field theory \cite{Hebeler:2013nza}.
This large number of EOSs is  employed  to evaluate the 
 properties  of  neutron star such as tidal deformability, radius and maximum mass. The correlations of neutron star properties 
with the pressure of $\beta$-equilibrated matter  at a given  density are studied.  Most  of  these correlations are  sensitive to the choice of the neutron star mass and EOS at a given density. Our results for the correlations of tidal deformability with pressure
for $\beta$-equilibrated matter are analogous  to those  obtained using a diverse set of nonrelativistic and relativistic mean-field models  (MFMs) that  re-emphasize  their model independence.
Such model-independent trends  inspire us  to parametrize  the pressure for $\beta$-equillibrated matter around 2$\rho_0$   in terms of  neutron star mass and the corresponding tidal deformability.

\subsection{Priors, likelihood and filters}

\begin{table}[hb]
\caption{\label{tab1} 
The prior distributions of the nuclear matter parameters . The nuclear matter parameters considered are the binding energy per nucleon ( $\varepsilon_0 $), incompressibility
coefficient ($K_0$), symmetry energy coefficient ($J_0$), it's slope parameter ($L_0$),
symmetry energy curvature parameter ($K_{\rm sym,0}$) and  $Q_0 (Q_{\rm sym,0})$ and $Z_0
(Z_{\rm sym,0})$ are related to third and fourth order density derivatives of  $E(\rho,0) $
( $E_{\rm sym}(\rho)$), respectively. All the nuclear matter parameters are evaluated at saturation
density $\rho_0$ =  0.16 fm$^{-3}$.  The parameters of  Gaussian  distribution (G) are the
mean ($\mu$) and standard deviation ($\sigma$).
} 
  \centering
  \begin{ruledtabular}  
  \begin{tabular}{cccc}
  %\toprule
NMPs &{Pr-Dist}&{$\mu$}&{$\sigma$}\\ [1.3ex]
(in MeV) &  &  & \\

\cline{1-4}
\hline 
{$\varepsilon_0$}   &  G & -16  & 0.3           \\[1.3ex] %\cline{1-7}

%\hline 
{$K_0$} & G &240  &  50            \\[1.3ex] %\cline{1-7}

%\hline 
{$Q_0$} &   G  & -400 & 400             \\[1.3ex] %\cline{1-7}

%\hline 
{$Z_0$} &   G   & 1500   & 1500            \\[1.3ex] %\cline{1-7}
  
%\hline
{$J_0$} &   G    &32  &5             \\[1.3ex] %\cline{1-7}
  
%\hline
{$L_0$} &    G   &50   &50           \\[1.3ex] %\cline{1-7}
  
%\hline

{$K_{\rm sym,0}$} &   G  &-100  &200              \\[1.3ex] %\cline{1-7}
%\hline

{$Q_{\rm sym,0}$} &   G   & 550&400             \\[1.3ex] %\cline{1-7}
  
%\hline

{$Z_{\rm sym,0}$} &   G   &-2000& 2000            \\[1.3ex] %\cline{1-7}
%\hline
  %\toprule
  \end{tabular}
  \end{ruledtabular}
\end{table}

We apply Bayesian approach to  obtain two large sets of EOSs
corresponding to the Taylor and $\frac{n}{3}$ expansions.  The posterior
distributions for the NMPs   are
obtained by subjecting the EOSs  to a  set of minimal constraints
which include some basic properties of nuclear matter evaluated
at the saturation density $\rho_0$  and  EOS for the 
pure neutron matter  at low-density. The constraints on the nuclear
matter parameters are incorporated through the priors and those
from the EOS for the  pure neutron matter through the likelihood
function. Not all the nuclear matter parameters are well constrained.
Only a very few  low-order nuclear matter parameters 
constrained within narrow bounds are the binding energy per nucleon
$\varepsilon_0=-16.0 \pm 0.3$ MeV , nuclear matter incompressibility
coefficients $K_0=240 \pm 50$ MeV for the symmetric
nuclear matter and symmetry energy coefficient $J_0=32.0 \pm 5 $
MeV. The values of $\varepsilon_0$ and $J_0$ are very well constrained by
the  binding energy of finite nuclei over a wide range of nuclear masses
\cite{Chabanat98,Chabanat97,Malik:2019whk,Mondal:2015tfa,Mondal:2016roo,Sulaksono:2009rn}.
The value of $K_0$ is constrained from the experimental data on the
centroid energy of isoscalar giant monopole resonance in a few heavy
nuclei \cite{Garg:2018uam,Agrawal:2005ix}.  The values of $L_0$ have been
extracted from experimental data   on variety of phenomena in the finite
nuclei as well as from neutron star observations. The model-independent
estimates of $L_0$ is expected to be  derived from the measurement
of neutron-skin thickness in asymmetric nuclei. Recent measurement
of neutron-skin thickness in $^{208}$Pb nucleus yields $L_0 = 106\pm
37$ MeV\cite{Reed:2021nqk}.  However, this value of $L_0$ has  only marginal overlap at the
lower side  with those determined using experimental data on iso-vector
giant dipole resonances in several nuclei\cite{Roca-Maza:2015eza} and recent neutron
star observations\cite{Essick:2021kjb}.  The remaining nuclear matter parameters,
$Q_0$, $Z_0$, $K_{\rm sym,0},
Q_{\rm sym,0}$ and $Z_{\rm sym,0}$  are constrained only poorly
\cite{Tsang:2020lmb,Ferreira:2021pni,Dutra:2012mb,Dutra:2014qga,Mondal:2017hnh}.
The priors for  the nuclear matter parameters employed in the present work
are listed in Table \ref{tab1}. The prior distributions of $\varepsilon_0,
K_0$ and  $J_0$ are assumed to be Gaussian with rather smaller width,
whereas, the other higher order nuclear matter parameters
  correspond to   Gaussian distribution with very
large width. We have also repeated our calculations  with uniform priors for the higher order nuclear matter parameters and the result  for the median values are found to be  practically unaltered and uncertainties are modified marginally, up to 10\%(not shown). In what follows, we present only those results which are obtained with priors as listed in Table \ref{tab1}.

We know that the direct application of the lattice QCD simulations
are challenging to hadronic physics at finite density due to sign
problem in Monte Carlo simulations. However, analytical calculations in
terms of the effective degrees of freedom at low energy ($\rho<\rho_0$)
like chiral effective theory is valid with negligible uncertainty. The
precise next-to-next-to-next-to-leading-order (N$^3$LO) calculation are
usually fitted to  the nucleon–deuteron scattering cross section or
few-body observables, and even saturation properties of heavier nuclei
\cite{Drischler:2021kxf}. The low-density EOS  for the pure neutron
matter obtained from  a (N$^{3}$LO) calculation in chiral effective field
theory \cite{Hebeler:2013nza} is employed as  pseudodata to obtain
a simple likelihood function as given by Eq. (\ref{eq:likelihood}).
The $d$s and the $\sigma$s  in Eq. (\ref{eq:likelihood}) are the pseudodata
for the energy per neutron and the  corresponding uncertainties taken from
Ref. \cite{Hebeler:2013nza}. This  has been employed in past many of the analyses as their pseudodata \cite{Ekstrom:2015rta,Lim:2018bkq,Lim:2019som,Malik:2022zol,Ghosh:2022lam}.
We have  considered the values of energy per neutron 
over the density range $\rho = 0.08$ - $0.16$fm$^{-3}$. At the densities
lower than $0.08$fm$^{-3}$, the neutron star matter  is expected to
be  clusterized.

We  have filtered  the nuclear matter parameters
 by demanding  that (i)  pressure for the
$\beta$-equilibrated matter  should  increase monotonically with density
(thermodynamic stability), (ii) speed of sound must not exceed the
speed of light (causality)  and (iii) maximum mass of neutron star must
exceeds $2M_\odot$ (observational constraint). The causality breaks down at higher density  mostly  for the  Taylor EOS.   In such cases, we use the stiffest EOS, $P(\epsilon)
= P_m + (\epsilon - \epsilon_m)$, where, $P_m$ and $\epsilon_m$ are the
pressure and corresponding energy density at which the causality breaks
\cite{Glendenning:1992dr}.

\subsection{ Posterior distribution of nuclear matter parameters}\label{PD}

To undertake the correlation systematics   as proposed, we need a large number of  EOSs with diverse behavior and corresponding neutron star  properties.    The posterior distributions for the nuclear matter parameters for the Taylor and $\frac{n}{3}$ expansions  are obtained by subjecting the EOS  to a  set of minimal constraints  as discussed above. The joint posterior distribution of the NMPs for a given model depends on the product of the likelihood and the prior distribution of  nuclear matter parameters (Eq. (\ref{eq:bt})).  The posterior distribution of each individual parameter is obtained by marginalizing the joint posterior distribution with the remaining model parameters. If the marginalized posterior distribution of a nuclear matter parameter is localized more than the corresponding prior distribution, then, the nuclear matter parameter is said to be well constrained by the data used for model fitting. 
   
%\newpage
\begin{figure*}
    \centering
    \includegraphics[width=1.05\textwidth,height=0.9\textheight]{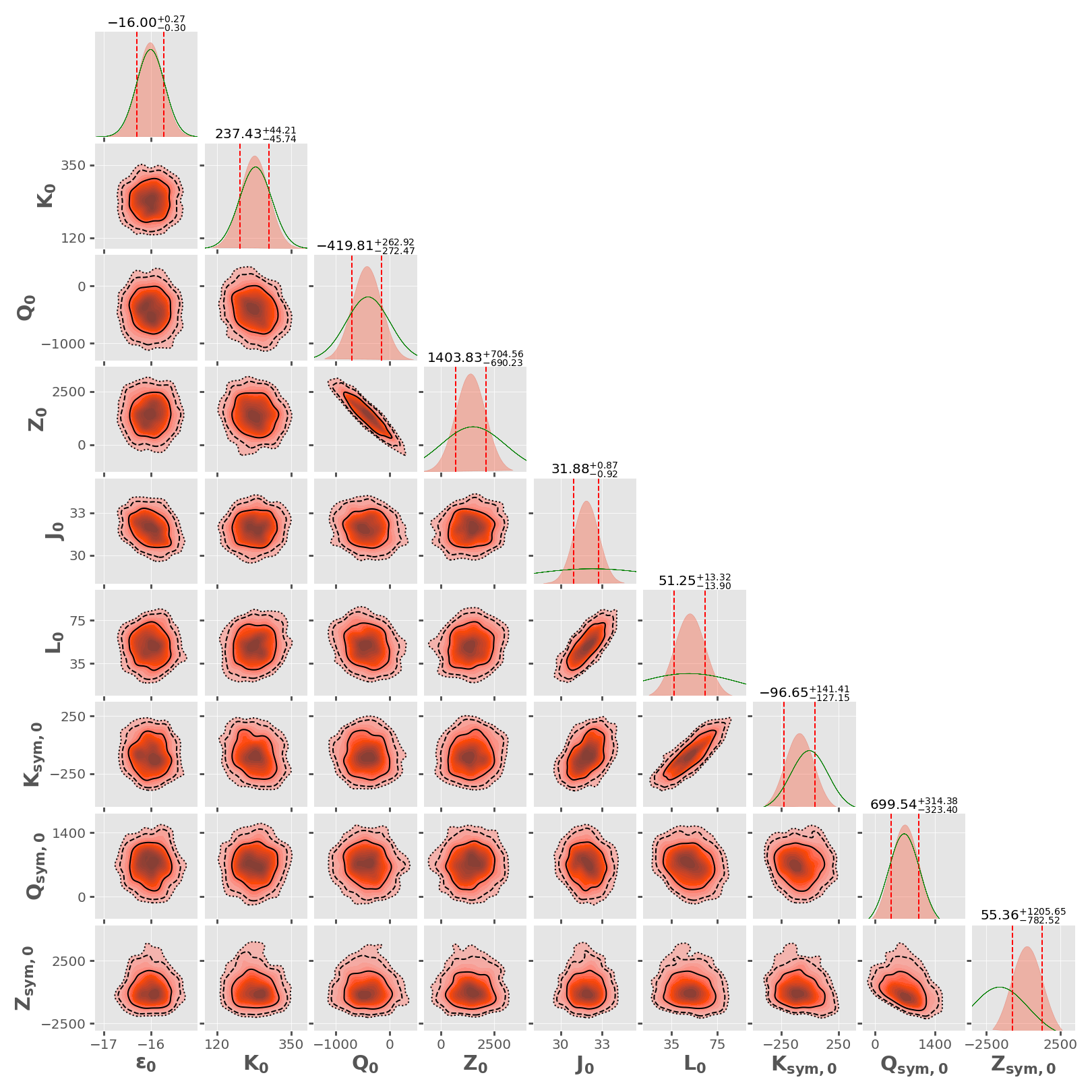}
       \caption{\label{fig1} Corner plots for the  nuclear matter parameters (in MeV) obtained for Taylor expansions for the EOS of asymmetric nuclear matter. The one dimensional marginalized posterior distributions (salmon) and the prior distributions (green lines) are displayed along the diagonal plots . The vertical lines indicate 68\% confidence interval of nuclear matter parameters. The confidence ellipses for two-dimensional posterior distributions are plotted with 1$\sigma$, 2$\sigma$ and 3$\sigma$ confidence intervals along the off-diagonal plots. The distributions of nuclear matter parameters are obtained by subjecting them to minimal constraints (see text for details). }
       
\end{figure*}
%\newpage
% \vspace{0.5mm}
 \begin{figure*}
    \centering
    \includegraphics[width=1.05\textwidth, height=0.9\textheight]{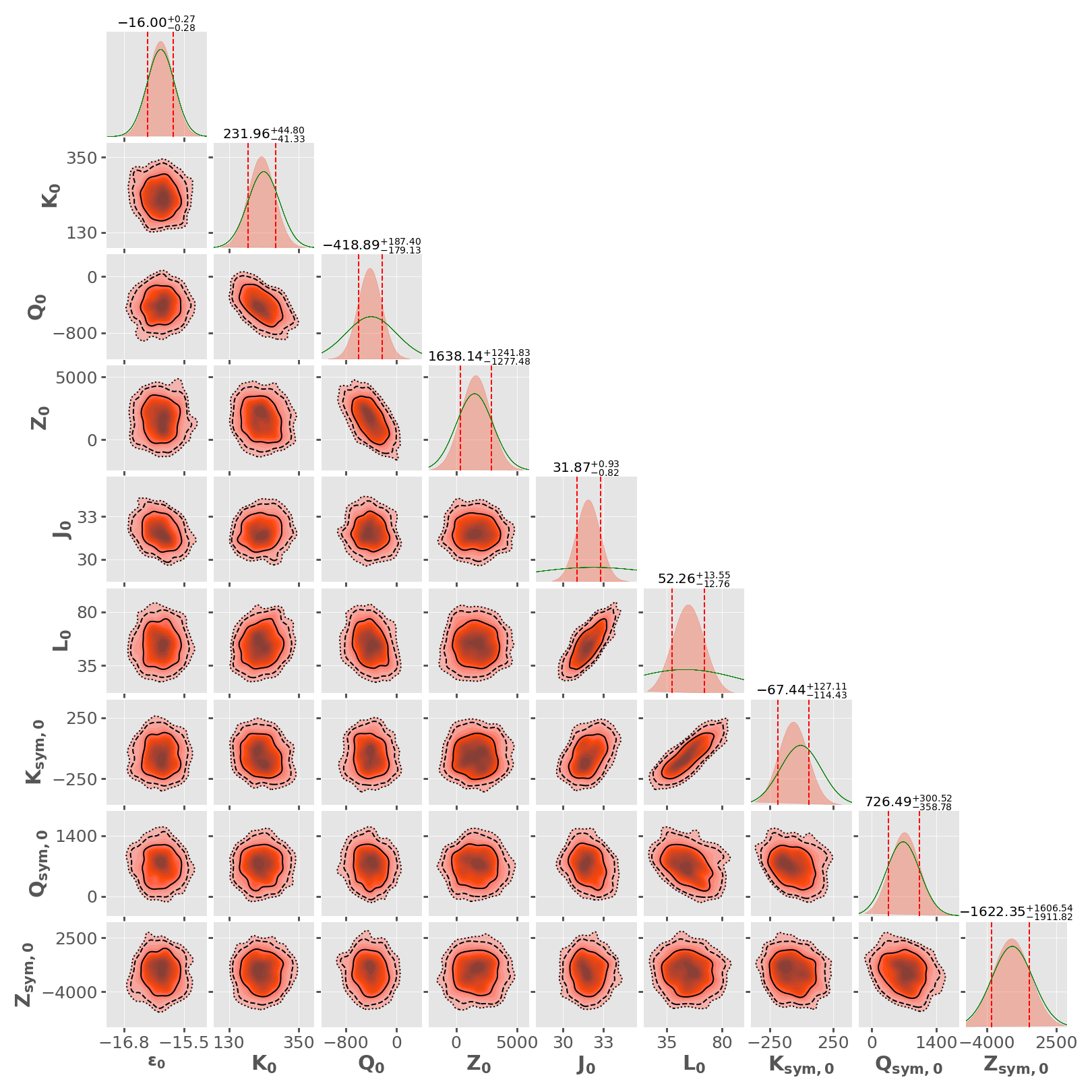} \caption{\label{fig2}
  The same as Fig. \ref{fig1}, but, for $\frac{n}{3}$  expansions for the EOS of asymmetric nuclear matter. }
\end{figure*}

The corner plots for the marginalized posterior distributions for the nuclear matter parameters in one and two dimensions  obtained for Taylor and $\frac{n}{3}$ expansions are displayed in Figs.\ref{fig1} and \ref{fig2}, respectively.
The differences between the one-dimensional posterior distributions for the nuclear matter parameters and corresponding prior distributions reflect the role of low-density EOS for pure neutron matter in constraining the nuclear matter parameters. The EOS for the pure neutron matter mainly constraints the values of $J_0$, $L_0$ and $K_{\rm sym,0}$ and to some extent $Q_{\rm sym,0}$ and $Z_{\rm sym,0}$. The shapes and the orientations of the confidence ellipses suggest that the correlations among most of the NMPs are weak. Most Strong correlations exist only between  $Q_0-Z_0$, $L_0-J_0$   and $L_0-K_{\rm sym,0}$  for both the expansions
with correlation coefficient r$\simeq$ 0.8. 
The $K_0 -Q_0$  correlation is  slightly better in case  of $\frac{n}{3}$ expansion (r$\sim$ -0.6) as compared to Taylor (r$\sim$ -0.18).
The median values of the nuclear matter parameters and the corresponding $68\% (90\%)$ confidence intervals obtained from the marginalized posterior distributions are listed in Table \ref{tab3}(see Appendix ). 
We also provide the values for the nuclear matter parameters obtained without the PNM constraints. The low-density pure neutron matter mainly constraints those nuclear matter parameters which are associated with the density dependence of the symmetry energy. The median values of $L_0$ and $K_{\rm sym, 0}$, which determined the linear and quadratic density dependence of the symmetry energy, become smaller suggesting softer symmetry energy "at high-density" with the inclusion of pure neutron matter constrains. Furthermore, the uncertainties on $L_0$ reduced by more than 50\%. The median value of $Q_{\rm sym,0}$ remain more or less unaltered. 
From the recent measurement 
of the neutron-skin thickness 
for $^{208}$Pb nucleus (PREX-II)\cite{PREX:2021umo,Reed:2021nqk},
 $\Delta R_{\rm skin}=0.283\pm0.071$ fm, 
 the value of
 $L_0$ has been determined to be $106\pm 37$ MeV \cite{Reed:2021nqk} . This value of $L_0$  agrees with the ones obtained in the
present work  with PNM constrained only  within $90\%$ confidence interval.

\subsection{Properties of neutron stars}

Once the EOS for the core and crust are known the values of NS mass, radius and tidal deformability corresponding to given central pressure can be obtained by solving Tolman-Oppenheimer-Volkoff equations \cite{Oppenheimer:1939ne,Tolman:1939jz}. The EOSs for core region of neutron star, correspond to the $\beta$-equilibrated matter over the density range $0.5-8\rho_0$, are obtained from  the   posterior distributions of nuclear matter parameters for the Taylor and $\frac{n}{3}$ expansions. The core EOSs are matched  to the crust EOSs for obtaining the NS properties. The EOS for outer crust is taken to be the  one given by Baym, Pethick, and Sutherland  \cite{Baym:1971pw}. The inner crust that joins the inner edge of the outer crust and the outer edge of the core is assumed to be polytropic \cite{Carriere:2002bx}, $p(\varepsilon)= c_1 + c_2 \varepsilon^{\gamma}$. Here, the parameters $c_1$ and $c_2$ are determined in such a way that the EOS for the inner crust matches with the outer crust at one end ($\rho = 10^{-4}$ fm$^{-3}$) and with the core at the other end ($0.5\rho_0$ ). The polytropic index $\gamma$ is taken to be equal to 4/3. The radii of neutron star with mass $\sim$ 1$M_\odot$ are more sensitive to the treatment of crust EOS  \cite{Fortin:2016hny}. It is demonstrated that the treatment of crust EOS employed in the present work may introduce the uncertainties of about 50-100 m in radii of NSs having mass 1.4$ M_\odot$. It is shown in Ref. \cite{Piekarewicz:2018sgy} that the choice of EOS for inner crust does not significantly impact the values of tidal deformability which depends on the Love number k$_{2}$ as well as the compactness parameter.

\begin{figure}[htp]
    \centering
    \includegraphics[width=0.48\textwidth]{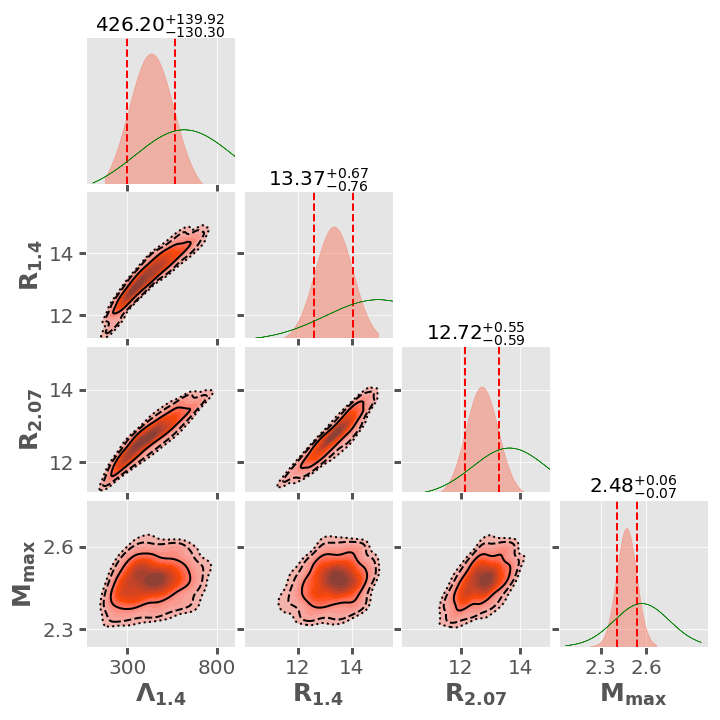}
    \vspace{1.0mm}
    \includegraphics[width=0.48\textwidth]{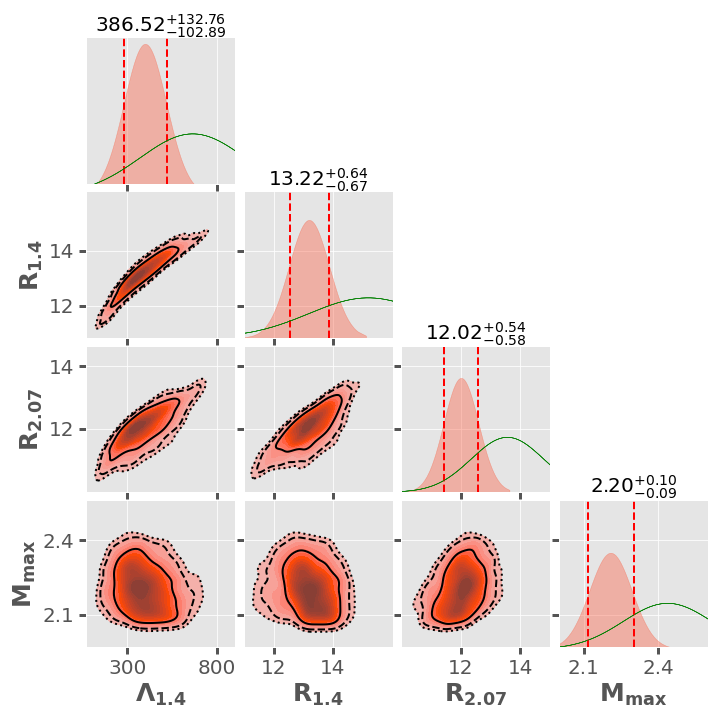}
    \caption{\label{fig3}Corner plots for the marginalized posterior distributions (salmon)  of
the tidal deformability $\Lambda_{1.4}$, radii $R_{1.4}$ (km) and $R_{2.07}$ (km)  and the maximum mass $M_{\rm max}$ ($M_\odot$) for Taylor (top) and $\frac{n}{3}$ (bottom) expansions.  The green lines  represent   effective  priors obtained using the priors
for nuclear matter parameters (see also Table \ref{tab1}).}
\end{figure}

We have obtained the distributions of $\Lambda_{1.4}$, $R_{1.4}$, $R_{2.07}$ and $M_{\rm max}$ using the  posterior distributions for the nuclear matter parameters corresponding to the Taylor and $\frac{n}{3}$ expansions. The corner plots for these NS properties are displayed in Fig.~\ref{fig3}. The effective priors for the NS properties as shown  by green lines are
obtained using the priors for the nuclear matter parameters. The posterior
distributions of NS properties  are  narrower than the corresponding
effective priors indicating the  significance of the low-density EOS for
the pure neutron matter. The posterior distributions of $\Lambda_{1.4}$ and $R_{1.4}$ for both the expansions
are quite close to each other. The differences begin to appear for the case of $R_{2.07}$ which become even  larger for the maximum mass.  This is due to the fact that the Taylor EOSs are  much more
stiffer than the those for $\frac{n}{3}$. The dichotomy in the high-density
behavior of the Taylor and $\frac{n}{3}$ expansions would help us to
understand the extent to which the correlations of the  EOSs with the
properties of NS, for  masses in the range  $1$ - $2M_\odot$, are
model dependent. It is clear 
 from off-diagonal plots that $\Lambda_{1.4}$ is strongly correlated with $R_{1.4}$, the correlation coefficient is  r$\sim$ 0.9. The $\Lambda_{1.4}$ and $R_{1.4}$ also display stronger correlations with $R_{2.07}$ (r $\sim$ 0.8) for the case of Taylor and somewhat moderate correlations (r $\sim$ 0.7) for the $\frac{n}{3}$ expansion. The maximum mass of neutron star is almost uncorrelated with the other NS properties considered.

\begin{figure}[hbp]
    \centering
    \includegraphics[width=0.5\textwidth]{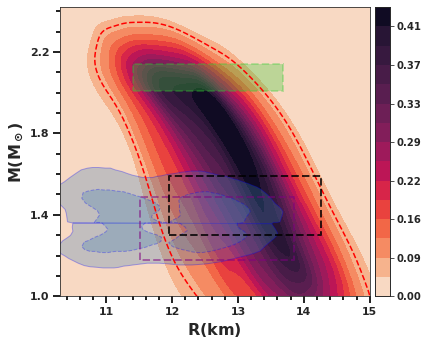}
    \caption{\label{fig4}
    Plot for joint probability distribution $P(M, R)$ as a function of  mass and radius of neutron star obtained for $\frac{n}{3}$ expansion. The red dashed line represents the 90\% confidence interval. The outer  and inner gray shaded regions indicate the 90\% (solid) and 50\% (dashed) confidence interval of the LIGO-Virgo analysis for BNS component from the GW170817 event \cite{GW170817_MR_PEsample,Abbott2019a,GWOSC_softx}. The rectangular regions enclosed by dotted lines indicate the constraints from the millisecond pulsar PSR J0030+0451 ( purple \& black ) NICER x-ray data \cite{Riley:2019yda,Miller:2019cac} and PSR J0740+6620 (green) \cite{Riley:2021pdl} }
\end{figure}

 We have  summarized in
 Table \ref{tab4} (see Appendix ) the  median values of NS properties along with 68\% (90\%) confidence intervals.   Like in the case of nuclear matter parameters, the NS properties get significantly constrained by the EOS of pure neutron matter at low-density. For instance, the median values of $\Lambda_{1.4}$ become smaller by about 15\% and the associated uncertainties by about 40\% with the pure neutron matter constraints. The median values of $R_{1.4}$ and the corresponding uncertainties also become noticeably smaller. 
The $R_{2.07}$ and $M_{\rm max}$ do not show significant changes with the inclusion of low-density pure neutron matter constraints. With the PNM constraints, the 90\% confidence interval of the neutron star properties such as tidal deformability, radius and mass overlap with the currently available bounds, $\Lambda_{1.4} \in [70,580]$ \cite{GW170817}, $R_{1.4} \in [11.41,13.61]$ km \cite{Miller:2021qha}, $R_{2.07} \in [11.8,13.1]$ km \cite{Riley:2021pdl} and $M_{\rm max} \geq 2.09 M_{\odot}$ \cite{Romani:2021xmb}.
The $M_{\rm max} = 2.48_{-0.07}^{+0.06} M_\odot$ obtained for the
Taylor EOSs  is on the slightly higher side in comparison to the ones
derived by combining the GW170817 observations of merging of binary
neutron stars and quasiuniversal relation  \cite{Rezzolla_2018}. The observed electromagnetic emissions in the form of kilonova and the detection of a gamma-ray burst has been linked to the formation of a black hole and thus, have been utilized to infer the maximum mass of a stable neutron star. However, such inference of the maximum mass is subjected to uncertainties originating from model dependence of postmerger dynamics. Recent observation of GW190814 event, a neutron star black hole/binary neutron star merger, has triggered an assessment of the maximum mass of a stable neutron star \cite{GW190814}. While there are different opinions available in the literature, the nature of a compact object in the range of 2.5 - 2.67 $M_\odot$ being neutron star or black hole seems to be an unsettled issue to date \cite{Tsokaros_2020,Lim_2021,Drischler_2021,Li:2021crp,Rezzolla_2018,GW190814}. So the maximum mass ($M_{max}$) we got for the Taylor model supporting the static NS of mass greater than 2.5 M$_\odot$ may not be ruled out at present.

We obtain joint probability distribution  $P(M,R)$  for a given mass and radius
for both the  Taylor and $\frac{n}{3}$  expansions. They display qualitatively very much similar  trends.  In Fig. \ref{fig4}, 
we plot the
$P(M,R)$ obtained for the  $\frac{n}{3}$ expansion.
The 90\% confidence interval is represented by  red dashed
line. The color gradient from orange to dark purple
represents the lowest to highest probability. The most probable values for $R_{1.4}$ and $R_{2.07}$ are  approximately 13.3 and 12.3 km, respectively. The $P(M,R)$ is maximum for $M\sim 
1.4 - 2.0 
M_{\odot}$, $R \sim 12.4 - 13.4
$ km. The $90\%$ confidence interval has 
significant
 overlap with LIGO-Virgo 
 and NICER estimations.  It may be however pointed out that the main objective of the present work is to construct large  sets of EOSs with diverse behavior  to assess various correlation systematics as follows.

\subsection{Correlations of neutron star properties with EOS}
\label{subsec_correl}
\begin{figure}[htbp]
\centering
%\hspace{-0.2cm}
\includegraphics[width=0.5\textwidth]{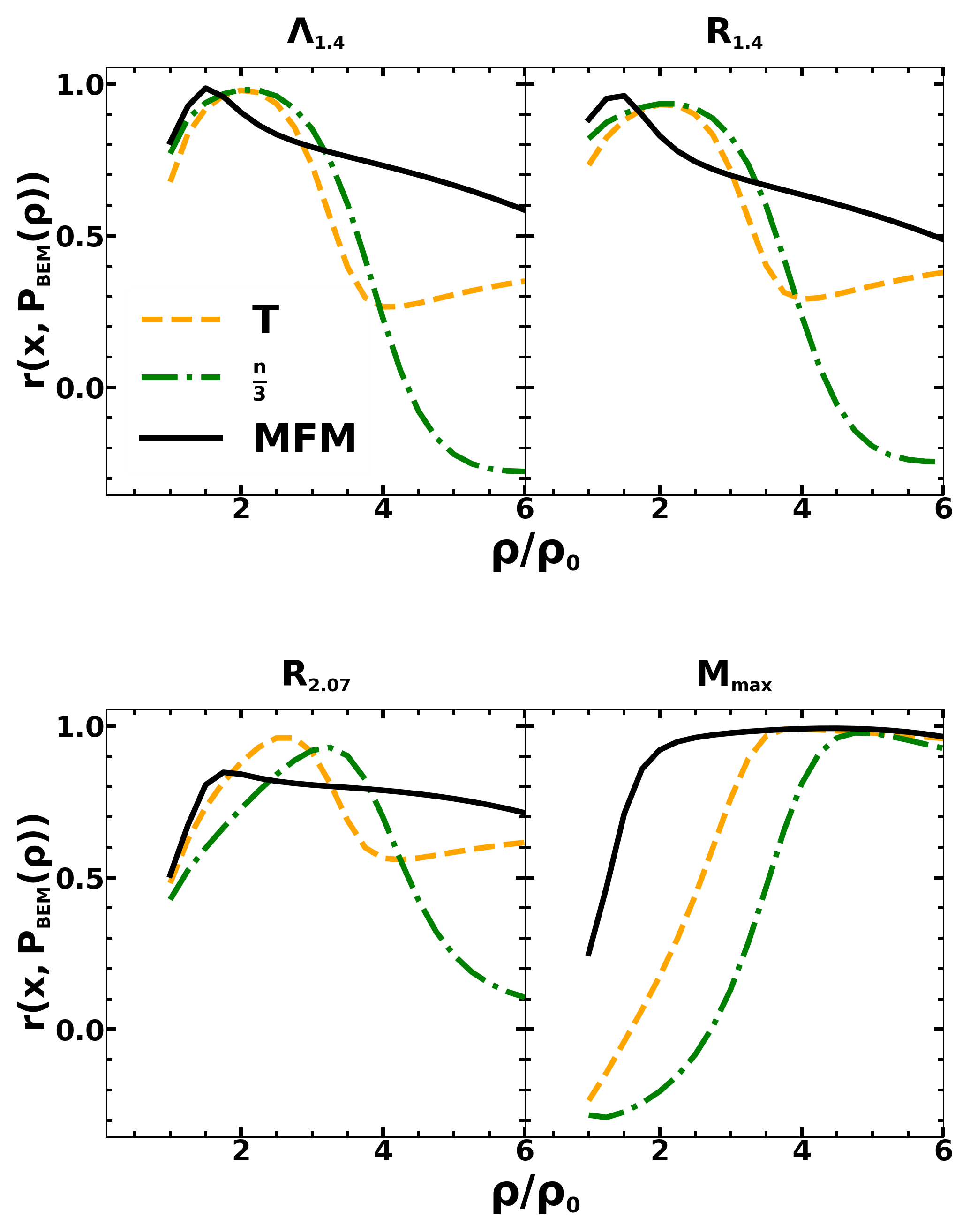}
\caption{\label{fig5}
The correlation coefficients r(x,$P_{\ms{\rm BEM}}(\rho)$), as approximated by both Taylor and $\frac{n}{3}$ expansion along with the mean-field theory calculations, is shown in this figure. Here, x represents either of the tidal deformability $\Lambda_{1.4}$, radii $R_{1.4}$, $R_{2.07}$, or maximum mass $M_{\rm max}$ of the neutron star, whereas, $P_{\ms{\rm BEM}}(\rho)$ represents the pressure for $\beta$-equilibrated matter  at a   density $\rho$. The calculations are performed with neutron star properties obtained  using  marginalized posterior distributions of nuclear matter parameters in Taylor and $\frac{n}{3}$ expansions. For the comparison the results are  also displayed for a diverse set of nonrelativistic and relativistic microscopic mean-field models (MFMs).}
\end{figure}

We randomly select 1000 EOSs and corresponding NS properties  from marginalized posterior distributions obtained for the Taylor as well as $\frac{n}{3}$ expansions. They are used to study the correlations of various NS properties with key quantities determining the behavior of the EOS. The correlations of $\Lambda_{1.4}$, $R_{1.4}$, $R_{2.07}$ and $M_{\rm max}$  with the pressure of $\beta$-equilibrated matter over a wide range of densities are evaluated. The values of correlation coefficients are plotted as a function of density in Fig.\ref{fig5}. We also display  the values of correlation coefficients for NS properties with the pressure of $\beta$-equilibrated matter calculated using unified EOSs for a diverse set of 41 nonrelativistic and relativistic microscopic mean-field models (MFMs) \cite{Fortin:2016hny}. The various NS properties considered show strong correlations with $P_{\ms{\rm BEM}}(\rho)$ around a particular density. The density at which the correlation is maximum increases with the NS mass. The values of $\Lambda_{1.4}$  and $R_{1.4}$ are strongly correlated with $P_{\ms{\rm BEM}}(\rho)$ at density $\sim$ $1.5-2.5\rho_0$. The $R_{2.07}$ is strongly correlated with $P_{\ms{\rm BEM}}(\rho)$ around 3$\rho_0$. The $M_{\rm max}$ is strongly correlated with $P_{\ms{\rm BEM}}(\rho)$  around 4.5$\rho_0$. Our results for the Taylor and $\frac{n}{3}$ expansions
 for the region of maximum correlations 
 are in line with those obtained using a diverse set of mean-field models, except for the case of $R_{2.07}$. Thus, it seems possible  that the EOS  over a range of densities beyond
$\rho_0$ can be constrained in a nearly model-independent  manner with
the help of various NS observables.

\begin{figure}[htbp]
\centering
%\hspace{-0.5cm}
\includegraphics[width=0.5\textwidth]{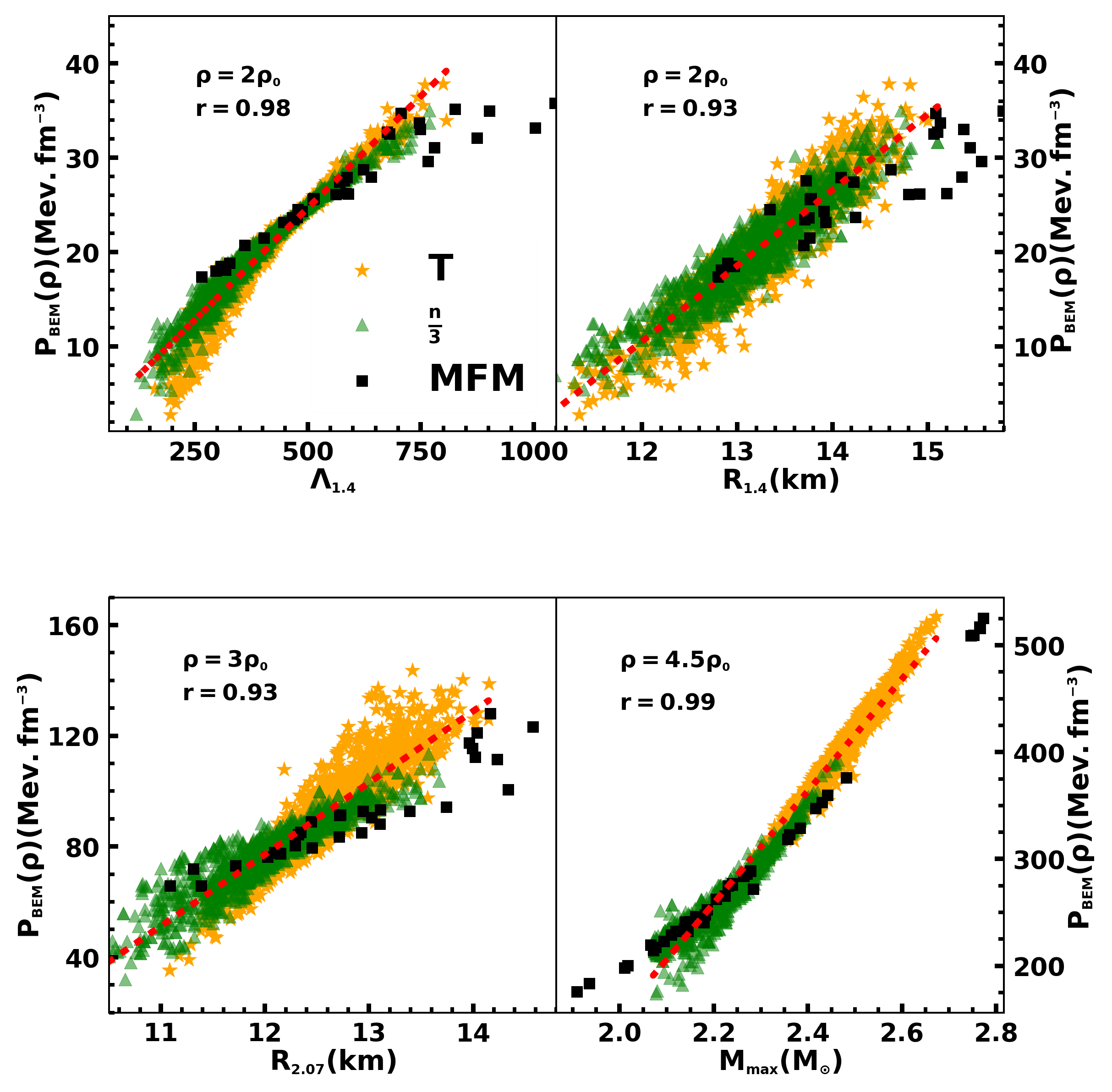}
\caption{\label{fig6}The variations of pressure for $\beta$-equilibrated matter [$P_{\ms{\rm BEM}}(\rho)$] at selected densities versus tidal deformability $\Lambda_{1.4}$, radii $R_{1.4}$ and $R_{2.07}$ and maximum mass $M_{\rm max}$ of neutron star. The red dashed lines are obtained by linear regression [see Eq.~(\ref{linear_regr} in Sec.\ref{subsec_correl}].}
\end{figure}

In Table \ref{tab5}(see the Appendix ), we list the values of correlation coefficients obtained between the NS properties and the EOS at some selected densities. The correlation coefficients are obtained using  100 and 1000 EOSs, corresponding to Taylor and $\frac{n}{3}$ expansions,
randomly selected from the posterior distributions. We also present the results  which are obtained by combining 1000 EOSs corresponding to each of the expansions. The values of correlation coefficients for the combined set of EOSs  are close to those obtained separately. The values of the correlation coefficients are close to those obtained for mean-field models are listed in second column.
We plot in Fig. \ref{fig6}, the variations of $P_{\ms{\rm BEM}}(\rho)$, at selected densities, with $\Lambda_{1.4}$, $R_{1.4}$, $R_{2.07}$ and $M_{\rm max}$ for which the correlations are stronger. We compare our results with  those obtained from a diverse set of mean-field models. The correlation lines  obtained by combining results of the Taylor and $\frac{n}{3}$ expansions are also plotted to estimate the values of $P_{\ms{\rm BEM}}(\rho)$ at selected densities with help of  NS properties. The equations for the correlation lines are obtained using linear regression as,
\bea
\label{linear_regr}
\frac{ P_{\ms{\rm BEM}}(2\rho_0)}{\rm MeV \rm fm^{-3}} &=& (0.96 \pm 0.10)   + (0.0473 \pm 0.0002)\Lambda_{1.4},\non
 \frac{ P_{\ms{\rm BEM}}(2\rho_0)}{\rm MeV \rm fm^{-3}} &=&  (-85.63 \pm 0.89) +  (8.01 \pm 0.06)\frac{ R_{1.4}}{\rm
km},\non
 \frac{ P_{\ms{\rm BEM}}(3\rho_0)}{\rm MeV \rm fm^{-3}} &=&
(-233.16 \pm 2.85) + (25.86 \pm 0.23) \frac{R_{2.07}}{ \rm km},\non
 \frac{ P_{\ms{\rm BEM}}(4.5\rho_0)}{\rm MeV \rm  fm^{-3}} &=&
 (-895.85 \pm 4.00) +  (524.75 \pm 1.70) \frac{M_{ \rm max}}{M_{\odot}}.\non
\eea

\begin{table*}[]
    %\centering
    \caption{\label{tab2} The median values and associated 68\%(90\%) uncertainties for the parameters, appearing in Eq. (\ref{eq_para}), obtained from their marginalized posterior distributions. The values of  parameters $b_0$, $b_1$ and $b_2$ as listed are scaled up by a factor of 10.}
   %  \begin{adjustwidth}{-2.2cm}{}
     %\small\addtolength{\tabcolsep}{-4.5pt}
     \begin{ruledtabular}
    \begin{tabular}{ccccccc}
    %\hline\hline
    Pressure & $a_0$ & $a_1$ & $a_2$ & $b_0$ & $b_1$ & $b_2$\\[1.5ex]
    ({in MeV fm$^{-3}$}) &  &   &  &  & & \\
    \hline
     $P_{\ms{\rm BEM}}(1.5\rho_0)$  & $0.544^{+0.031(0.050)}_{-0.029(0.060)}$ & $1.869^{+0.158(0.260)}_{-0.161(0.309)}$ & $7.451^{+0.237(0.390)}_{-0.234(0.450)}$ & $0.176^{+0.001(0.001)}_{-0.001(0.002)}$ & $0.740^{+0.004(0.007)}_{-0.004(0.008)}$ &  $1.152^{+0.013(0.021)}_{-0.012(0.025)}$ \\[1.6ex]
     $P_{\ms{\rm BEM}}(2\rho_0)$ &  $0.146^{+0.030(0.050)}_{-0.030(0.059)}$  & $-0.598^{+0.163(0.269)}_{-0.159(0.320)}$ & $27.909^{+0.233(0.397)}_{-0.240(0.469)}$ & $0.493^{+0.001(0.001)}_{-0.001(0.002)}$  & $2.234^{+0.004(0.007)}_{-0.004(0.008)}$ & $3.728^{+0.012(0.021)}_{-0.012(0.025)}$ \\[1.6ex]
     $P_{\ms{\rm BEM}}(2.5\rho_0)$ & $7.345^{+0.030(0.050)}_{-0.030(0.060)}$  & $-15.102^{+0.161(0.272)}_{-0.167(0.321)}$ & $68.411^{+0.239(0.396)}_{-0.238(0.475)}$ & $0.906^{+0.001(0.001)}_{-0.001(0.002)}$ &  $4.518^{+0.004(0.007)}_{-0.004(0.008)}$ & $8.115^{+0.012(0.021)}_{-0.012(0.025)}$ \\[1.6ex]
     %\hline\hline
  
    \end{tabular}
    \end{ruledtabular}
    %\end{adjustwidth}
\end{table*}

\begin{figure}[htbp]
    \centering
    \includegraphics[width=0.5\textwidth]{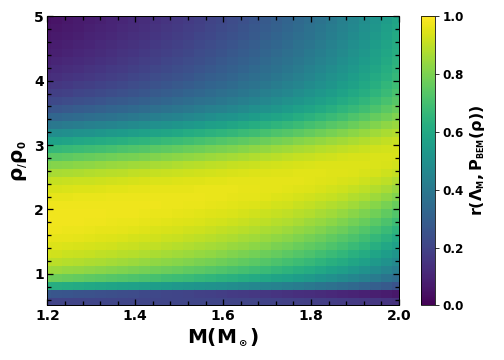}
    \caption{\label{fig7} Dependence of correlation coefficients between tidal deformability ($\Lambda_M$)  and the pressure of $\beta$-equilibrated matter ($P_{\ms{\rm BEM}}(\rho)$) on neutron star mass (M) and density ($\rho$) is depicted in this plot. Here $\rho_0$=0.16 fm$^{-3}$ is used only for scaling purposes. }
\end{figure}

\begin{figure}[]
    \centering
    %\hspace{-0.5cm}
    \includegraphics[width=0.5\textwidth]{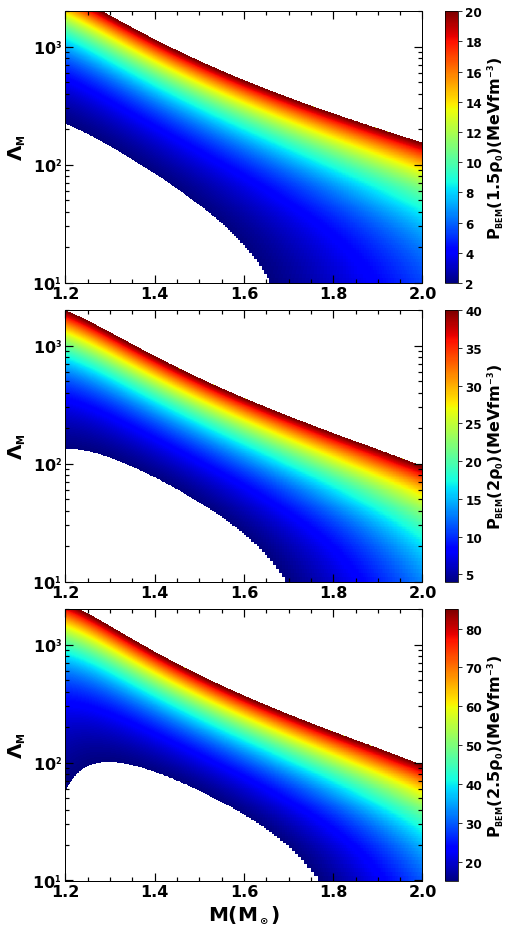}
    \caption{\label{fig8} The median values of pressure for $\beta$-equilibrated matter is shown here as a function of neutron star mass and its tidal deformability at densities 1.5$\rho_0$ (top), 2.0$\rho_0$ (middle) and 2.5$\rho_0$ (bottom).}
    
\end{figure}

We extend our analysis for the correlations of the pressure for the $\beta$-equilibrated matter with tidal deformability over a wide range of neutron stars mass. In Fig. \ref{fig7}, we display color-coded graph for the correlations of tidal deformability of neutron star for the mass $1.2-2.0 M_\odot$ with the pressure for $\beta$-equilibrated matter at densities $0.5-5\rho_0$ (r($\Lambda_M,P_{\ms{\rm BEM}}(\rho)$)). One can easily obtain the value of correlation coefficient as a function of density at a given NS mass. The $P_{\ms{\rm BEM}}(\rho)$ at $\rho \sim 1.5-2.5\rho_0$ are strongly correlated ($r \sim 0.8-1$ ) with tidal deformability for NS masses  in the range $1.2-2.0 M_\odot$. Hence,
$P_{\ms{\rm BEM}} (\rho)$ can be parametrized at a given $\rho$ as,
\bea
\frac{ P_{\ms{\rm BEM}}(\rho)}{\rm MeV \rm fm^{-3}}
 &=&a(M) +b(M) \Lambda_M, \label{eq_para}
\eea
with  mass-dependent coefficients $a(M)$ and $b(M)$ expanded as
\bea
a(M) &=&  (a_0 + a_1(M-M_0) + a_2(M-M_0)^2), \\
b(M) &=&  ( b_0 + b_1(M-M_0) + b_2(M-M_0)^2), \label{eq_para2}
\eea
respectively, where $M_0$ is taken to be 1.4$M_\odot$ and the values of $a_i$ and $b_i$ are estimated using a Bayesian approach with the help of  $P_{\ms{\rm BEM}} (\rho)$ and tidal deformability obtained for Taylor and $\frac{n}{3}$ expansions. For a given $\rho$, the Eq. (\ref{eq_para}) is fitted using the tidal deformability corresponding to NS mass $1.2-2.0 M_\odot$. The priors for $a_i$ and $b_i$ are taken to be uniform in the range of -100 to 100. The calculations are performed for $\rho$= 1.5, 2.0 and 2.5 $\rho_0$. All the $a_i$s are strongly correlated with corresponding $b_i$s.
The median values of parameters $a_i$ and $b_i$ and associated  uncertainties are summarized in Table \ref{tab2}. It may be noticed that the values of $a_0$ and $b_0$ for the case of
$P_{\ms{\rm BEM}}  (2\rho_0)$ are not the same as those in Eq. (\ref{linear_regr}).  This may be partly due to the  strong correlations between 
 $a_0$ and $b_0$
of Eq. (\ref{eq_para}). Moreover, Eq. (\ref{linear_regr}) is fitted to the  values of tidal deformability at a fixed NS mass 1.4$M_\odot$. 
To validate our  parametrized form for $P_{\ms{\rm BEM}}(\rho)$, we have calculated the
values of  $P_{\ms{\rm BEM}}(2\rho_0)$ using Eq. (\ref{eq_para}) with the help of tidal
deformability for 1.4$M_\odot$ obtained for large number of mean-field
models which includes the once considered  in Fig.  \ref{fig5} along
with those taken from \cite{Tsang:2019vxn,Malik:2022zol,Ferreira:2019bgy}.
The  average deviation of  $P_{\ms{\rm BEM}}(2 \rho_0)$, obtained using
Eq. (\ref{eq_para}),  from the actual values is  about  $10\%$.
We find marginal improvement when the terms corresponding to quadratic in tidal deformability are included in Eq. (\ref{eq_para}). 

%\newpage
In Fig.\ref{fig8}, we display the variations of tidal deformability as a function of mass and pressure for $\beta$-equilibrated matter at $\rho$=1.5, 2.0 and 2.5 $\rho_0$. These results are obtained using the parametrized form for  $P_{\ms{\rm BEM}}(\rho)$ as given by Eq. (\ref{eq_para}). One can easily estimate the values of $P_{\ms{\rm BEM}}(\rho)$ for $\rho \sim 2\rho_0$ once the values of tidal deformability known in NS mass ranges $1.2-2.0 M_\odot$.

%\newpage
\section{Conclusions} \label{summary}
We have used Taylor and $\frac{n}{3}$ expansions of equations of state to construct marginalized posterior
distributions of the nuclear matter parameters which are consistent with the minimal constraints. Only a few low-order nuclear matter parameters,  such as the energy per
nucleon, incompressibility  coefficient for the symmetric nuclear matter and symmetry energy coefficients at the saturation density ($\rho_0$), are constrained in  narrow windows along with the  the low-density pure neutron matter EOS obtained from a precise next-to-next-to-next-to-leading-order (N$^{3}$LO) calculation in chiral effective field theory.  The tidal deformability, radius and maximum mass are evaluated using large sets of minimally constrained EOSs.

The correlations of neutron star properties over a wide range of mass with various key quantities characterizing the EOS are investigated. We find that the values of tidal  deformability and radius for the neutron star with 1.4$M_\odot$  are strongly correlated with  the pressure for the $\beta$-equilibrated matter at density
$\sim 2\rho_0$. The radius for $2.07M_\odot$ neutron star is strongly correlated with the pressure for $\beta$-equilibrated matter at density $\sim 3\rho_0$.   The maximum mass of neutron star is  correlated with the pressure for the $\beta$-equilibrated matter at density   $\sim 4.5\rho_0$.  These correlation systematics
are in harmony with those  obtained for unified EOSs for the $\beta$-equilibrated matter available for  a diverse set of nonrelativistic and relativistic mean-field models. We exploit the  model independence of correlations to   parametrize the pressure for $\beta$-equilibrated matter, in the density range $1.5-2.5\rho_0$, in terms of the mass and corresponding tidal deformability of neutron star. Such  parametric form may facilitate back-of-the-envelope  estimation of  the  pressure at densities around $2\rho_0$ for a given value of tidal deformability of neutron stars with mass in the range of $1.2-2.0 M_{\odot}$.

 \section{Acknowledgements} 
The authors would like to thank C. Mondal, for a careful reading of the paper, and important suggestions. 
N.K.P. would like to thank T.K. Jha for constant encouragement and support and gratefully acknowledge the Department of Science and Technology, Ministry of Science and Technology, India, for the support of DST/INSPIRE Fellowship/2019/IF190058. AM acknowledges support from the DST-SERB Start-up Research Grant SRG/2020/001290. The authors sincerely acknowledge the usage of the analysis software {\tt BILBY} \cite{Ashton2019,Bilby_ref} and open data from GWOSC \cite{GWOSC_softx}. This research has made use of data or software obtained from the gravitational-wave Open Science Center (gw-openscience.org), a service of LIGO Laboratory, the LIGO Scientific Collaboration, the Virgo Collaboration, and KAGRA. LIGO Laboratory and Advanced LIGO are funded by the United States National Science Foundation (NSF) as well as the Science and Technology Facilities Council (STFC) of the United Kingdom, the Max-Planck-Society (MPS), and the State of Niedersachsen/Germany for support of the construction of Advanced LIGO and construction and operation of the GEO600 detector. Additional support for Advanced LIGO was provided by the Australian Research Council. Virgo is funded, through the European Gravitational Observatory (EGO), by the French Centre National de Recherche Scientifique (CNRS), the Italian Istituto Nazionale di Fisica Nucleare (INFN) and the Dutch Nikhef, with contributions by institutions from Belgium, Germany, Greece, Hungary, Ireland, Japan, Monaco, Poland, Portugal, Spain. The construction and operation of KAGRA are funded by Ministry of Education, Culture, Sports, Science and Technology (MEXT), and Japan Society for the Promotion of Science (JSPS), National Research Foundation (NRF) and Ministry of Science and ICT (MSIT) in Korea, Academia Sinica (AS) and the Ministry of Science and Technology (MoST) in Taiwan. TM acknowledges support from FCT (Fundação para a Ciência e a Tecnologia, I.P, Portugal) under the Projects No. UID/\-FIS/\-04564/\-2019, No. UIDP/\-04564/\-2020, No. UIDB/\-04564/\-2020, and No. POCI-01-0145-FEDER-029912 with financial support from Science, Technology and Innovation, in its FEDER component, and by the FCT/MCTES budget through national funds (OE).

 \begin{figure*}[]
    \begin{minipage}{\textwidth}
    \centering .
    \appendix
   \section{ Some useful tables}%\label{appendix}
 	\end{minipage}

We present our results in tabular form which are obtained with the
minimal constraints.  The values of the nuclear matters, properties
of neutron stars  and their correlations with various  key quantities
associated with EOS  are listed in Tables III–V. These results
are depicted in Figs. \ref{fig1} - \ref{fig6}.
\end{figure*}                                             
%\captionsetup{labelformat=AppendixTables}
\setcounter{table}{2} 

\begin{table*}[]
    \centering
    \caption{\label{tab3} The median values and associated 68\%(90\%) uncertainties for the nuclear matter parameters  from their marginalized posterior distributions. The results are obtained for  Taylor and $\frac{n}{3}$ expansions with and without pure neutron matter (PNM) constraints.}
    \begin{ruledtabular} 
    \begin{tabular}{ccccc}
    %\toprule
    
  \multirow{2}{*}{NMPs} & \multicolumn{2}{c}{ without PNM}  & \multicolumn{2}{c}{with PNM} \\ [1.5ex] 
  \cline{2-5}
    {(in MeV)}    & Taylor & $\frac{n}{3}$ & Taylor & $\frac{n}{3}$ \\[1.5ex] 
    \cline{1-5}    
   % \hline\hline
    $\varepsilon_0$ & $-16.02^{+0.23(0.41)}_{-0.28(0.56)}$  &$-15.99^{+0.27(0.43)}_{-0.27(0.51)}$  &$-16.00^{+0.27(0.42)}_{-0.30(0.54)}$  & $-16.00^{+0.27(0.44)}_{-0.28(0.56)}$  \\[1.5ex] 
    $K_0$ & $236.42^{+42.78(74.34)}_{-42.58(79.62)}$  &$233.38^{+48.94(76.14)}_{-42.73(83.95)}$ & $237.43^{+44.24(72.25)}_{-45.75(83.22)}$ & $231.96^{+44.80(72.94)}_{-41.33(76.63)}$  \\ [1.5ex] 
    $Q_0$ & $-436.23^{+273.36(419.17)}_{-306.50(603.76)}$ & $-411.84^{+207.53(301.56)}_{-210.88(409.00)}$  & $-419.81^{+262.95(437.69)}_{-272.47(531.58)}$  & $-418.89^{+187.43(300.76)}_{-179.25(377.42)}$   \\ [1.5ex]  
    $Z_0$ & $1441.51^{+792.45(1298.64)}_{-696.39(1381.30)}$ & $1600.07^{+1067.33(1883.00)}_{-1362.28(2615.10)}$  & $1403.84^{+704.56(1133.85)}_{-690.82(1386.25)}$  & $1638.14^{+1241.83(1906.75)}_{-1277.48(2244.23)}$ \\ [1.5ex]  
    $J_0$ & $32.37^{+4.08(6.79)}_{-4.26(8.83)}$ & $32.37^{+4.69(7.22)}_{-4.71(10.23)}$  & $31.88^{+0.87(1.43)}_{-0.92(-1.85)}$  & $31.87^{+0.93(1.49)}_{-0.82(1.68)}$   \\ [1.5ex]  
   $L_0$ & $59.88^{+41.14(65.90)}_{-39.84(78.17)}$ & $55.60^{+37.59(63.89)}_{-43.88(84.62)}$  & $51.25^{+13.32(21.60)}_{-13.91(25.54)}$ & $52.25^{+13.55(22.73)}_{-12.76(23.04)}$  \\ [1.5ex]  
    $K_{\rm sym,0}$ & $-85.86^{+192.67(327.83)}_{-151.57(266.76)}$  & $-40.03^{+161.60(271.89)}_{-135.08(234.67)}$  &
     $-96.65^{+141.41(225.69)}_{-127.49(216.74)}$ & $-67.44^{+127.18(206.09)}_{-114.80(200.38)}$   \\ [1.5ex] 
    $Q_{\rm sym,0}$ & $731.13^{+308.54(543.01)}_{-347.82(669.47)}$  & $705.36^{+311.23(511.39)}_{-352.72(727.86)}$ & $699.56^{+324.38(521.95)}_{-323.52(639.30)}$ & $726.49^{+300.40(510.33)}_{-358.51(631.86)}$  \\ [1.5ex] 
    $Z_{\rm sym,0}$ & $-2.07^{+1190.67(2153.84)}_{-820.92(1473.09)}$ &
    $-1390.39^{+1518.69(2526.53)}_{-1856.18(3623.74)}$  & $55.34^{+1205.62(2255.28)}_{-782.52(1415.84)}$  & $-1622.35^{+1606.61(2788.70)}_{-1911.81(3468.40)}$    \\ [1.5ex] 
   
    %\hline 
    %\toprule
    \end{tabular}
    
    \end{ruledtabular}
    %\end{adjustwidth}
\end{table*}

\begin{table*}[]
    \centering
    \caption{\label{tab4} Similar to Table \ref{tab3}, but, for the neutron star properties, namely  the tidal deformability ($\Lambda_{1.4}$), radii ($R_{1.4}$ and $R_{2.07}$)  and maximum mass ($M_{\rm max}$) .}
   % \begin{adjustwidth}{-0.5cm}{}
   % \small\addtolength{\tabcolsep}{-1pt}
    \begin{ruledtabular} 
    \begin{tabular}{ccccc}
   % \hline\hline
  \multirow{2}{*}{{NS properties}} & \multicolumn{2}{c}{ without PNM}  & \multicolumn{2}{c}{with PNM} \\ 
  \cline{2-5}
        & Taylor & $\frac{n}{3}$ & Taylor & $\frac{n}{3}$ \\[1.5ex] 
    \cline{1-5}   
   $\Lambda_{1.4}$ & $527.72^{+250.72(477.68)}_{-186.11(292.57)}$  & $455.85^{+223.65(465.72)}_{-163.05(243.23)}$ & $426.20^{+139.93(224.58)}_{-130.32(205.18)}$ & $386.52^{+132.76(213.24)}_{-102.84(199.09)}$  \\[1.5ex]
   $R_{1.4}${(km)} & $14.69^{+1.78(3.43)}_{-1.63(2.74)}$  & $14.15^{+1.87(3.34)}_{-1.69(2.58)}$  & $13.37^{+0.67(1.03)}_{-0.75(1.60)}$  & $13.22^{+0.64(0.99)}_{-0.67(1.59)}$   \\[1.5ex]
   $R_{2.07}${(km)} & $13.24^{+0.82(1.49)}_{-0.82(1.42)}$  & $12.27^{+0.88(1.52)}_{-0.80(1.52)}$  & $12.72^{+0.55(0.85)}_{-0.59(1.07)}$  & $12.02^{+0.54(0.88)}_{-0.58(1.23)}$  \\[1.5ex]
  $ M_{\rm max}${($M_\odot$)} & $2.45^{+0.07(0.11)}_{-0.06(0.13)}$  & $2.19^{+0.10(0.19)}_{-0.09(0.09)}$ & $2.48^{+0.06(0.10)}_{-0.07(0.14)}$  & $2.20^{+0.10(0.16)}_{-0.09(0.11)}$   \\[1.5ex]
%\hline\hline 
    \end{tabular}
    \end{ruledtabular}
   % \end{adjustwidth}
\end{table*}

\begin{table*}[]
    \centering
    \caption{\label{tab5} The comparison of values for Pearson's correlation coefficient (r) obtained from randomly selected 100 and 1000
    EOSs using both Taylor and $\frac{n}{3}$ expansions. The values of correlation coefficients are also obtained by combining  1000
    EOSs from each of the expansions.
    For comparison,  the values of $r$ obtained for a diverse set of mean-field
models are also presented in 2nd column.    }
     \begin{ruledtabular} 
    \begin{tabular}{ccccccc}
      %  \hline\hline
        \multirow{2}{*}{Name of Pairs}& MFMs & \multicolumn{2}{c}{Taylor} & \multicolumn{2}{c}{$\frac{n}{3}$} & combined\\
         & 41 & 100 & 1000 & 100 & 1000  &  2000\\
         \hline
        $\Lambda_{1.4}$-$P_{\ms{\rm BEM}}(2\rho_0)$ & 0.90 &0.98 & 0.98 & 0.99 & 0.98 & 0.98 \\[1.5ex]
        $R_{1.4}$-$P_{\ms{\rm BEM}}(2\rho_0)$ & 0.83 &0.93 & 0.93 & 0.94 & 0.93 & 0.93\\[1.5ex]
        $R_{2.07}$-$P_{\ms{\rm BEM}}(3\rho_0)$ & 0.81 &0.93 & 0.91 & 0.92 & 0.92 & 0.93\\[1.5ex]
        $M_{\rm max}$-$P_{\ms{\rm BEM}}(4.5\rho_0)$ & 0.99 &0.97 & 0.98 & 0.95 & 0.96 & 0.99\\[1.5ex]
       % \hline\hline
    \end{tabular}
    
    \end{ruledtabular}
\end{table*}

\clearpage
\newpage


\begin{thebibliography}{106}%
\makeatletter
\providecommand \@ifxundefined [1]{%
 \@ifx{#1\undefined}
}%
\providecommand \@ifnum [1]{%
 \ifnum #1\expandafter \@firstoftwo
 \else \expandafter \@secondoftwo
 \fi
}%
\providecommand \@ifx [1]{%
 \ifx #1\expandafter \@firstoftwo
 \else \expandafter \@secondoftwo
 \fi
}%
\providecommand \natexlab [1]{#1}%
\providecommand \enquote  [1]{``#1''}%
\providecommand \bibnamefont  [1]{#1}%
\providecommand \bibfnamefont [1]{#1}%
\providecommand \citenamefont [1]{#1}%
\providecommand \href@noop [0]{\@secondoftwo}%
\providecommand \href [0]{\begingroup \@sanitize@url \@href}%
\providecommand \@href[1]{\@@startlink{#1}\@@href}%
\providecommand \@@href[1]{\endgroup#1\@@endlink}%
\providecommand \@sanitize@url [0]{\catcode `\\12\catcode `\$12\catcode
  `\&12\catcode `\#12\catcode `\^12\catcode `\_12\catcode `\%12\relax}%
\providecommand \@@startlink[1]{}%
\providecommand \@@endlink[0]{}%
\providecommand \url  [0]{\begingroup\@sanitize@url \@url }%
\providecommand \@url [1]{\endgroup\@href {#1}{\urlprefix }}%
\providecommand \urlprefix  [0]{URL }%
\providecommand \Eprint [0]{\href }%
\providecommand \doibase [0]{http://dx.doi.org/}%
\providecommand \selectlanguage [0]{\@gobble}%
\providecommand \bibinfo  [0]{\@secondoftwo}%
\providecommand \bibfield  [0]{\@secondoftwo}%
\providecommand \translation [1]{[#1]}%
\providecommand \BibitemOpen [0]{}%
\providecommand \bibitemStop [0]{}%
\providecommand \bibitemNoStop [0]{.\EOS\space}%
\providecommand \EOS [0]{\spacefactor3000\relax}%
\providecommand \BibitemShut  [1]{\csname bibitem#1\endcsname}%
\let\auto@bib@innerbib\@empty
%</preamble>
\bibitem [{\citenamefont {Abbott~{\sl et
  al.}}(2019{\natexlab{a}})}]{Abbott18a}%
  \BibitemOpen
  \bibfield  {author} {\bibinfo {author} {\bibfnamefont {B.~P.}\ \bibnamefont
  {Abbott~{\sl et al.}}},\ }\href@noop {} {\bibfield  {journal} {\bibinfo
  {journal} {Phys. Rev. X}\ }\textbf {\bibinfo {volume} {9}},\ \bibinfo {pages}
  {011001} (\bibinfo {year} {2019}{\natexlab{a}})}\BibitemShut {NoStop}%
\bibitem [{\citenamefont {Abbott~{\sl et
  al.}}(2019{\natexlab{b}})}]{Abbott2019}%
  \BibitemOpen
  \bibfield  {author} {\bibinfo {author} {\bibfnamefont {B.~P.}\ \bibnamefont
  {Abbott~{\sl et al.}}},\ }\href@noop {} {\bibfield  {journal} {\bibinfo
  {journal} {Astrophys. J.}\ }\textbf {\bibinfo {volume} {882}},\ \bibinfo
  {pages} {L24} (\bibinfo {year} {2019}{\natexlab{b}})}\BibitemShut {NoStop}%
\bibitem [{\citenamefont {Abbott~{\sl et
  al.}}(2020{\natexlab{a}})}]{Abbott2020}%
  \BibitemOpen
  \bibfield  {author} {\bibinfo {author} {\bibfnamefont {B.~P.}\ \bibnamefont
  {Abbott~{\sl et al.}}},\ }\href@noop {} {\bibfield  {journal} {\bibinfo
  {journal} {Astrophys. J.}\ }\textbf {\bibinfo {volume} {892}},\ \bibinfo
  {pages} {L3} (\bibinfo {year} {2020}{\natexlab{a}})}\BibitemShut {NoStop}%
\bibitem [{\citenamefont {Abbott~{\sl et al.}}(2017)}]{GW170817}%
  \BibitemOpen
  \bibfield  {author} {\bibinfo {author} {\bibfnamefont {B.~P.}\ \bibnamefont
  {Abbott~{\sl et al.}}},\ }\href@noop {} {\bibfield  {journal} {\bibinfo
  {journal} {Phys. Rev. Lett.}\ }\textbf {\bibinfo {volume} {119}},\ \bibinfo
  {pages} {161101} (\bibinfo {year} {2017})}\BibitemShut {NoStop}%
\bibitem [{\citenamefont {Malik}\ \emph {et~al.}(2018)\citenamefont {Malik},
  \citenamefont {Alam}, \citenamefont {Fortin}, \citenamefont
  {Provid{\^{e}}ncia}, \citenamefont {Agrawal}, \citenamefont {Jha},
  \citenamefont {Kumar},\ and\ \citenamefont {Patra}}]{Malik2018}%
  \BibitemOpen
  \bibfield  {author} {\bibinfo {author} {\bibfnamefont {T.}~\bibnamefont
  {Malik}}, \bibinfo {author} {\bibfnamefont {N.}~\bibnamefont {Alam}},
  \bibinfo {author} {\bibfnamefont {M.}~\bibnamefont {Fortin}}, \bibinfo
  {author} {\bibfnamefont {C.}~\bibnamefont {Provid{\^{e}}ncia}}, \bibinfo
  {author} {\bibfnamefont {B.~K.}\ \bibnamefont {Agrawal}}, \bibinfo {author}
  {\bibfnamefont {T.~K.}\ \bibnamefont {Jha}}, \bibinfo {author} {\bibfnamefont
  {B.}~\bibnamefont {Kumar}}, \ and\ \bibinfo {author} {\bibfnamefont {S.~K.}\
  \bibnamefont {Patra}},\ }\href@noop {} {\bibfield  {journal} {\bibinfo
  {journal} {Phys. Rev. C}\ }\textbf {\bibinfo {volume} {98}},\ \bibinfo
  {pages} {035804} (\bibinfo {year} {2018})}\BibitemShut {NoStop}%
\bibitem [{\citenamefont {De}\ \emph {et~al.}(2018)\citenamefont {De},
  \citenamefont {Finstad}, \citenamefont {Lattimer}, \citenamefont {Brown},
  \citenamefont {Berger},\ and\ \citenamefont {Biwer}}]{De18}%
  \BibitemOpen
  \bibfield  {author} {\bibinfo {author} {\bibfnamefont {S.}~\bibnamefont
  {De}}, \bibinfo {author} {\bibfnamefont {D.}~\bibnamefont {Finstad}},
  \bibinfo {author} {\bibfnamefont {J.~M.}\ \bibnamefont {Lattimer}}, \bibinfo
  {author} {\bibfnamefont {D.~A.}\ \bibnamefont {Brown}}, \bibinfo {author}
  {\bibfnamefont {E.}~\bibnamefont {Berger}}, \ and\ \bibinfo {author}
  {\bibfnamefont {C.~M.}\ \bibnamefont {Biwer}},\ }\href@noop {} {\bibfield
  {journal} {\bibinfo  {journal} {Phys. Rev. Lett.}\ }\textbf {\bibinfo
  {volume} {121}},\ \bibinfo {pages} {091102} (\bibinfo {year}
  {2018})}\BibitemShut {NoStop}%
\bibitem [{\citenamefont {Fattoyev}\ \emph {et~al.}(2018)\citenamefont
  {Fattoyev}, \citenamefont {Piekarewicz},\ and\ \citenamefont
  {Horowitz}}]{Fattoyev2018a}%
  \BibitemOpen
  \bibfield  {author} {\bibinfo {author} {\bibfnamefont {F.~J.}\ \bibnamefont
  {Fattoyev}}, \bibinfo {author} {\bibfnamefont {J.}~\bibnamefont
  {Piekarewicz}}, \ and\ \bibinfo {author} {\bibfnamefont {C.~J.}\ \bibnamefont
  {Horowitz}},\ }\href@noop {} {\bibfield  {journal} {\bibinfo  {journal}
  {Phys. Rev. Lett.}\ }\textbf {\bibinfo {volume} {120}},\ \bibinfo {pages}
  {172702} (\bibinfo {year} {2018})}\BibitemShut {NoStop}%
\bibitem [{\citenamefont {Landry}\ and\ \citenamefont
  {Essick}(2019)}]{Landry2019}%
  \BibitemOpen
  \bibfield  {author} {\bibinfo {author} {\bibfnamefont {P.}~\bibnamefont
  {Landry}}\ and\ \bibinfo {author} {\bibfnamefont {R.}~\bibnamefont
  {Essick}},\ }\href@noop {} {\bibfield  {journal} {\bibinfo  {journal} {Phys.
  Rev. D}\ }\textbf {\bibinfo {volume} {99}},\ \bibinfo {pages} {084049}
  (\bibinfo {year} {2019})}\BibitemShut {NoStop}%
\bibitem [{\citenamefont {Piekarewicz}\ and\ \citenamefont
  {Fattoyev}(2019{\natexlab{a}})}]{Piekarewicz2019}%
  \BibitemOpen
  \bibfield  {author} {\bibinfo {author} {\bibfnamefont {J.}~\bibnamefont
  {Piekarewicz}}\ and\ \bibinfo {author} {\bibfnamefont {F.~J.}\ \bibnamefont
  {Fattoyev}},\ }\href@noop {} {\bibfield  {journal} {\bibinfo  {journal}
  {Phys. Rev. C}\ }\textbf {\bibinfo {volume} {99}},\ \bibinfo {pages} {045802}
  (\bibinfo {year} {2019}{\natexlab{a}})}\BibitemShut {NoStop}%
\bibitem [{\citenamefont {Malik}\ \emph {et~al.}(2019)\citenamefont {Malik},
  \citenamefont {Agrawal}, \citenamefont {De}, \citenamefont {Samaddar},
  \citenamefont {Provid{\^{e}}ncia}, \citenamefont {Mondal},\ and\
  \citenamefont {Jha}}]{Malik:2019whk}%
  \BibitemOpen
  \bibfield  {author} {\bibinfo {author} {\bibfnamefont {T.}~\bibnamefont
  {Malik}}, \bibinfo {author} {\bibfnamefont {B.~K.}\ \bibnamefont {Agrawal}},
  \bibinfo {author} {\bibfnamefont {J.~N.}\ \bibnamefont {De}}, \bibinfo
  {author} {\bibfnamefont {S.~K.}\ \bibnamefont {Samaddar}}, \bibinfo {author}
  {\bibfnamefont {C.}~\bibnamefont {Provid{\^{e}}ncia}}, \bibinfo {author}
  {\bibfnamefont {C.}~\bibnamefont {Mondal}}, \ and\ \bibinfo {author}
  {\bibfnamefont {T.~K.}\ \bibnamefont {Jha}},\ }\href@noop {} {\bibfield
  {journal} {\bibinfo  {journal} {Phys. Rev. C}\ }\textbf {\bibinfo {volume}
  {99}},\ \bibinfo {pages} {052801} (\bibinfo {year} {2019})}\BibitemShut
  {NoStop}%
\bibitem [{\citenamefont {Biswas}\ \emph {et~al.}(2021)\citenamefont {Biswas},
  \citenamefont {Char}, \citenamefont {Nandi},\ and\ \citenamefont
  {Bose}}]{Biswas2020}%
  \BibitemOpen
  \bibfield  {author} {\bibinfo {author} {\bibfnamefont {B.}~\bibnamefont
  {Biswas}}, \bibinfo {author} {\bibfnamefont {P.}~\bibnamefont {Char}},
  \bibinfo {author} {\bibfnamefont {R.}~\bibnamefont {Nandi}}, \ and\ \bibinfo
  {author} {\bibfnamefont {S.}~\bibnamefont {Bose}},\ }\href {\doibase
  10.1103/PhysRevD.103.103015} {\bibfield  {journal} {\bibinfo  {journal}
  {Phys. Rev. D}\ }\textbf {\bibinfo {volume} {103}},\ \bibinfo {pages}
  {103015} (\bibinfo {year} {2021})}\BibitemShut {NoStop}%
\bibitem [{\citenamefont {Thi}\ \emph {et~al.}(2021{\natexlab{a}})\citenamefont
  {Thi}, \citenamefont {Mondal},\ and\ \citenamefont {Gulminelli}}]{Thi2021}%
  \BibitemOpen
  \bibfield  {author} {\bibinfo {author} {\bibfnamefont {H.~D.}\ \bibnamefont
  {Thi}}, \bibinfo {author} {\bibfnamefont {C.}~\bibnamefont {Mondal}}, \ and\
  \bibinfo {author} {\bibfnamefont {F.}~\bibnamefont {Gulminelli}},\
  }\href@noop {} {\bibfield  {journal} {\bibinfo  {journal} {Universe}\
  }\textbf {\bibinfo {volume} {7}},\ \bibinfo {pages} {373} (\bibinfo {year}
  {2021}{\natexlab{a}})}\BibitemShut {NoStop}%
\bibitem [{\citenamefont {Watts~{\sl et al.}}(2016)}]{Watts2016}%
  \BibitemOpen
  \bibfield  {author} {\bibinfo {author} {\bibfnamefont {A.~L.}\ \bibnamefont
  {Watts~{\sl et al.}}},\ }\href@noop {} {\bibfield  {journal} {\bibinfo
  {journal} {Rev. Mod. Phys.}\ }\textbf {\bibinfo {volume} {88}},\ \bibinfo
  {pages} {021001} (\bibinfo {year} {2016})}\BibitemShut {NoStop}%
\bibitem [{\citenamefont {Miller~{\sl et
  al.}}(2019{\natexlab{a}})}]{Miller2019}%
  \BibitemOpen
  \bibfield  {author} {\bibinfo {author} {\bibfnamefont {M.~C.}\ \bibnamefont
  {Miller~{\sl et al.}}},\ }\href@noop {} {\bibfield  {journal} {\bibinfo
  {journal} {Astrophys. J.}\ }\textbf {\bibinfo {volume} {887}},\ \bibinfo
  {pages} {L24} (\bibinfo {year} {2019}{\natexlab{a}})}\BibitemShut {NoStop}%
\bibitem [{\citenamefont {Riley~{\sl et al.}}(2019{\natexlab{a}})}]{Riley2019}%
  \BibitemOpen
  \bibfield  {author} {\bibinfo {author} {\bibfnamefont {T.}~\bibnamefont
  {Riley~{\sl et al.}}},\ }\href@noop {} {\bibfield  {journal} {\bibinfo
  {journal} {Astrophys. J. Lett.}\ }\textbf {\bibinfo {volume} {887}},\
  \bibinfo {pages} {L21} (\bibinfo {year} {2019}{\natexlab{a}})}\BibitemShut
  {NoStop}%
\bibitem [{\citenamefont {Zhang}\ \emph {et~al.}(2018)\citenamefont {Zhang},
  \citenamefont {Li},\ and\ \citenamefont {Xu}}]{Zhang2018}%
  \BibitemOpen
  \bibfield  {author} {\bibinfo {author} {\bibfnamefont {N.-B.}\ \bibnamefont
  {Zhang}}, \bibinfo {author} {\bibfnamefont {B.-A.}\ \bibnamefont {Li}}, \
  and\ \bibinfo {author} {\bibfnamefont {J.}~\bibnamefont {Xu}},\ }\href@noop
  {} {\bibfield  {journal} {\bibinfo  {journal} {Astrophys. J.}\ }\textbf
  {\bibinfo {volume} {859}},\ \bibinfo {pages} {90} (\bibinfo {year}
  {2018})}\BibitemShut {NoStop}%
\bibitem [{\citenamefont {Cai}\ and\ \citenamefont {Li}(2021)}]{Cai2021}%
  \BibitemOpen
  \bibfield  {author} {\bibinfo {author} {\bibfnamefont {B.~J.}\ \bibnamefont
  {Cai}}\ and\ \bibinfo {author} {\bibfnamefont {B.~A.}\ \bibnamefont {Li}},\
  }\href@noop {} {\bibfield  {journal} {\bibinfo  {journal} {Phys. Rev. C}\
  }\textbf {\bibinfo {volume} {103}},\ \bibinfo {pages} {034607} (\bibinfo
  {year} {2021})}\BibitemShut {NoStop}%
\bibitem [{\citenamefont {Gil}\ \emph {et~al.}(2022)\citenamefont {Gil},
  \citenamefont {Papakonstantinou},\ and\ \citenamefont {Hyun}}]{Gil:2021ols}%
  \BibitemOpen
  \bibfield  {author} {\bibinfo {author} {\bibfnamefont {H.}~\bibnamefont
  {Gil}}, \bibinfo {author} {\bibfnamefont {P.}~\bibnamefont
  {Papakonstantinou}}, \ and\ \bibinfo {author} {\bibfnamefont {C.~H.}\
  \bibnamefont {Hyun}},\ }\href@noop {} {\bibfield  {journal} {\bibinfo
  {journal} {Int. J. Mod. Phys. E}\ }\textbf {\bibinfo {volume} {31}},\
  \bibinfo {pages} {2250013} (\bibinfo {year} {2022})}\BibitemShut {NoStop}%
\bibitem [{\citenamefont {Lattimer}\ and\ \citenamefont
  {Prakash}(2001)}]{Lattimer:2000nx}%
  \BibitemOpen
  \bibfield  {author} {\bibinfo {author} {\bibfnamefont {J.~M.}\ \bibnamefont
  {Lattimer}}\ and\ \bibinfo {author} {\bibfnamefont {M.}~\bibnamefont
  {Prakash}},\ }\href@noop {} {\bibfield  {journal} {\bibinfo  {journal}
  {Astrophys. J.}\ }\textbf {\bibinfo {volume} {550}},\ \bibinfo {pages} {426}
  (\bibinfo {year} {2001})}\BibitemShut {NoStop}%
\bibitem [{\citenamefont {Maselli}\ \emph {et~al.}(2013)\citenamefont
  {Maselli}, \citenamefont {Cardoso}, \citenamefont {Ferrari}, \citenamefont
  {Gualtieri},\ and\ \citenamefont {Pani}}]{Maselli:2013mva}%
  \BibitemOpen
  \bibfield  {author} {\bibinfo {author} {\bibfnamefont {A.}~\bibnamefont
  {Maselli}}, \bibinfo {author} {\bibfnamefont {V.}~\bibnamefont {Cardoso}},
  \bibinfo {author} {\bibfnamefont {V.}~\bibnamefont {Ferrari}}, \bibinfo
  {author} {\bibfnamefont {L.}~\bibnamefont {Gualtieri}}, \ and\ \bibinfo
  {author} {\bibfnamefont {P.}~\bibnamefont {Pani}},\ }\href@noop {} {\bibfield
   {journal} {\bibinfo  {journal} {Phys. Rev. D}\ }\textbf {\bibinfo {volume}
  {88}},\ \bibinfo {pages} {023007} (\bibinfo {year} {2013})}\BibitemShut
  {NoStop}%
\bibitem [{\citenamefont {Abbott~{\sl et
  al.}}(2018{\natexlab{a}})}]{Abbot2018}%
  \BibitemOpen
  \bibfield  {author} {\bibinfo {author} {\bibfnamefont {B.~P.}\ \bibnamefont
  {Abbott~{\sl et al.}}},\ }\href@noop {} {\bibfield  {journal} {\bibinfo
  {journal} {Phys. Rev. Lett.}\ }\textbf {\bibinfo {volume} {121}},\ \bibinfo
  {pages} {161101} (\bibinfo {year} {2018}{\natexlab{a}})}\BibitemShut
  {NoStop}%
\bibitem [{\citenamefont {Lim}\ and\ \citenamefont {Holt}(2018)}]{Lim:2018bkq}%
  \BibitemOpen
  \bibfield  {author} {\bibinfo {author} {\bibfnamefont {Y.}~\bibnamefont
  {Lim}}\ and\ \bibinfo {author} {\bibfnamefont {J.~W.}\ \bibnamefont {Holt}},\
  }\href@noop {} {\bibfield  {journal} {\bibinfo  {journal} {Phys. Rev. Lett.}\
  }\textbf {\bibinfo {volume} {121}},\ \bibinfo {pages} {062701} (\bibinfo
  {year} {2018})}\BibitemShut {NoStop}%
\bibitem [{\citenamefont {Tsang}\ \emph {et~al.}(2019)\citenamefont {Tsang},
  \citenamefont {Tsang}, \citenamefont {Danielewicz}, \citenamefont {Lynch},\
  and\ \citenamefont {Fattoyev}}]{Tsang:2019vxn}%
  \BibitemOpen
  \bibfield  {author} {\bibinfo {author} {\bibfnamefont {C.~Y.}\ \bibnamefont
  {Tsang}}, \bibinfo {author} {\bibfnamefont {M.~B.}\ \bibnamefont {Tsang}},
  \bibinfo {author} {\bibfnamefont {P.}~\bibnamefont {Danielewicz}}, \bibinfo
  {author} {\bibfnamefont {W.~G.}\ \bibnamefont {Lynch}}, \ and\ \bibinfo
  {author} {\bibfnamefont {F.~J.}\ \bibnamefont {Fattoyev}},\ }\href@noop {}
  {\bibfield  {journal} {\bibinfo  {journal} {Phys. Lett. B}\ }\textbf
  {\bibinfo {volume} {796}},\ \bibinfo {pages} {1} (\bibinfo {year}
  {2019})}\BibitemShut {NoStop}%
\bibitem [{\citenamefont {Tsang}\ \emph {et~al.}(2020)\citenamefont {Tsang},
  \citenamefont {Tsang}, \citenamefont {Danielewicz}, \citenamefont {Lynch},\
  and\ \citenamefont {Fattoyev}}]{Tsang:2020lmb}%
  \BibitemOpen
  \bibfield  {author} {\bibinfo {author} {\bibfnamefont {C.~Y.}\ \bibnamefont
  {Tsang}}, \bibinfo {author} {\bibfnamefont {M.~B.}\ \bibnamefont {Tsang}},
  \bibinfo {author} {\bibfnamefont {P.}~\bibnamefont {Danielewicz}}, \bibinfo
  {author} {\bibfnamefont {W.~G.}\ \bibnamefont {Lynch}}, \ and\ \bibinfo
  {author} {\bibfnamefont {F.~J.}\ \bibnamefont {Fattoyev}},\ }\href@noop {}
  {\bibfield  {journal} {\bibinfo  {journal} {Phys. Rev. C}\ }\textbf {\bibinfo
  {volume} {102}},\ \bibinfo {pages} {045808} (\bibinfo {year}
  {2020})}\BibitemShut {NoStop}%
\bibitem [{\citenamefont {Ashton~{\sl et al.}}(2019)}]{Ashton2019}%
  \BibitemOpen
  \bibfield  {author} {\bibinfo {author} {\bibfnamefont {G.}~\bibnamefont
  {Ashton~{\sl et al.}}},\ }\href@noop {} {\bibfield  {journal} {\bibinfo
  {journal} {Astrophys. J. Suppl. Ser.}\ }\textbf {\bibinfo {volume} {241}},\
  \bibinfo {pages} {27} (\bibinfo {year} {2019})}\BibitemShut {NoStop}%
\bibitem [{\citenamefont {Biscoveanu}\ \emph {et~al.}(2020)\citenamefont
  {Biscoveanu}, \citenamefont {Thrane},\ and\ \citenamefont
  {Vitale}}]{Biscoveanu2020a}%
  \BibitemOpen
  \bibfield  {author} {\bibinfo {author} {\bibfnamefont {S.}~\bibnamefont
  {Biscoveanu}}, \bibinfo {author} {\bibfnamefont {E.}~\bibnamefont {Thrane}},
  \ and\ \bibinfo {author} {\bibfnamefont {S.}~\bibnamefont {Vitale}},\
  }\href@noop {} {\bibfield  {journal} {\bibinfo  {journal} {Astrophys. J.}\
  }\textbf {\bibinfo {volume} {893}},\ \bibinfo {pages} {38} (\bibinfo {year}
  {2020})}\BibitemShut {NoStop}%
\bibitem [{\citenamefont {Coughlin}\ and\ \citenamefont
  {Dietrich}(2019)}]{Coughlin2019}%
  \BibitemOpen
  \bibfield  {author} {\bibinfo {author} {\bibfnamefont {M.~W.}\ \bibnamefont
  {Coughlin}}\ and\ \bibinfo {author} {\bibfnamefont {T.}~\bibnamefont
  {Dietrich}},\ }\href@noop {} {\bibfield  {journal} {\bibinfo  {journal}
  {Phys. Rev. D}\ }\textbf {\bibinfo {volume} {100}},\ \bibinfo {pages}
  {043011} (\bibinfo {year} {2019})}\BibitemShut {NoStop}%
\bibitem [{\citenamefont {{Hernandez Vivanco}}\ \emph
  {et~al.}(2019)\citenamefont {{Hernandez Vivanco}}, \citenamefont {Smith},
  \citenamefont {Thrane}, \citenamefont {Lasky}, \citenamefont {Talbot},\ and\
  \citenamefont {Raymond}}]{HernandezVivanco2019}%
  \BibitemOpen
  \bibfield  {author} {\bibinfo {author} {\bibfnamefont {F.}~\bibnamefont
  {{Hernandez Vivanco}}}, \bibinfo {author} {\bibfnamefont {R.}~\bibnamefont
  {Smith}}, \bibinfo {author} {\bibfnamefont {E.}~\bibnamefont {Thrane}},
  \bibinfo {author} {\bibfnamefont {P.~D.}\ \bibnamefont {Lasky}}, \bibinfo
  {author} {\bibfnamefont {C.}~\bibnamefont {Talbot}}, \ and\ \bibinfo {author}
  {\bibfnamefont {V.}~\bibnamefont {Raymond}},\ }\href@noop {} {\bibfield
  {journal} {\bibinfo  {journal} {Phys. Rev. D}\ }\textbf {\bibinfo {volume}
  {100}},\ \bibinfo {pages} {103009} (\bibinfo {year} {2019})}\BibitemShut
  {NoStop}%
\bibitem [{\citenamefont {Biscoveanu}\ \emph {et~al.}(2019)\citenamefont
  {Biscoveanu}, \citenamefont {Vitale},\ and\ \citenamefont
  {Haster}}]{Biscoveanu2019a}%
  \BibitemOpen
  \bibfield  {author} {\bibinfo {author} {\bibfnamefont {S.}~\bibnamefont
  {Biscoveanu}}, \bibinfo {author} {\bibfnamefont {S.}~\bibnamefont {Vitale}},
  \ and\ \bibinfo {author} {\bibfnamefont {C.-J.}\ \bibnamefont {Haster}},\
  }\href@noop {} {\bibfield  {journal} {\bibinfo  {journal} {Astrophys. J.}\
  }\textbf {\bibinfo {volume} {884}},\ \bibinfo {pages} {L32} (\bibinfo {year}
  {2019})}\BibitemShut {NoStop}%
\bibitem [{\citenamefont {Lower}\ \emph {et~al.}(2018)\citenamefont {Lower},
  \citenamefont {Thrane}, \citenamefont {Lasky},\ and\ \citenamefont
  {Smith}}]{Lower2018}%
  \BibitemOpen
  \bibfield  {author} {\bibinfo {author} {\bibfnamefont {M.~E.}\ \bibnamefont
  {Lower}}, \bibinfo {author} {\bibfnamefont {E.}~\bibnamefont {Thrane}},
  \bibinfo {author} {\bibfnamefont {P.~D.}\ \bibnamefont {Lasky}}, \ and\
  \bibinfo {author} {\bibfnamefont {R.}~\bibnamefont {Smith}},\ }\href@noop {}
  {\bibfield  {journal} {\bibinfo  {journal} {Phys. Rev. D}\ }\textbf {\bibinfo
  {volume} {98}},\ \bibinfo {pages} {083028} (\bibinfo {year}
  {2018})}\BibitemShut {NoStop}%
\bibitem [{\citenamefont {Romero-Shaw}\ \emph {et~al.}(2019)\citenamefont
  {Romero-Shaw}, \citenamefont {Lasky},\ and\ \citenamefont
  {Thrane}}]{Romero-Shaw2019}%
  \BibitemOpen
  \bibfield  {author} {\bibinfo {author} {\bibfnamefont {I.~M.}\ \bibnamefont
  {Romero-Shaw}}, \bibinfo {author} {\bibfnamefont {P.~D.}\ \bibnamefont
  {Lasky}}, \ and\ \bibinfo {author} {\bibfnamefont {E.}~\bibnamefont
  {Thrane}},\ }\href@noop {} {\bibfield  {journal} {\bibinfo  {journal} {Mon.
  Not. R. Astron. Soc.}\ }\textbf {\bibinfo {volume} {490}},\ \bibinfo {pages}
  {5210} (\bibinfo {year} {2019})}\BibitemShut {NoStop}%
\bibitem [{\citenamefont {Ramos-Buades}\ \emph {et~al.}(2020)\citenamefont
  {Ramos-Buades}, \citenamefont {Husa}, \citenamefont {Pratten}, \citenamefont
  {Estell{\'{e}}s}, \citenamefont {Garc{\'{i}}a-Quir{\'{o}}s}, \citenamefont
  {Mateu-Lucena}, \citenamefont {Colleoni},\ and\ \citenamefont
  {Jaume}}]{Ramos-Buades2020}%
  \BibitemOpen
  \bibfield  {author} {\bibinfo {author} {\bibfnamefont {A.}~\bibnamefont
  {Ramos-Buades}}, \bibinfo {author} {\bibfnamefont {S.}~\bibnamefont {Husa}},
  \bibinfo {author} {\bibfnamefont {G.}~\bibnamefont {Pratten}}, \bibinfo
  {author} {\bibfnamefont {H.}~\bibnamefont {Estell{\'{e}}s}}, \bibinfo
  {author} {\bibfnamefont {C.}~\bibnamefont {Garc{\'{i}}a-Quir{\'{o}}s}},
  \bibinfo {author} {\bibfnamefont {M.}~\bibnamefont {Mateu-Lucena}}, \bibinfo
  {author} {\bibfnamefont {M.}~\bibnamefont {Colleoni}}, \ and\ \bibinfo
  {author} {\bibfnamefont {R.}~\bibnamefont {Jaume}},\ }\href@noop {}
  {\bibfield  {journal} {\bibinfo  {journal} {Phys. Rev. D}\ }\textbf {\bibinfo
  {volume} {101}},\ \bibinfo {pages} {083015} (\bibinfo {year}
  {2020})}\BibitemShut {NoStop}%
\bibitem [{\citenamefont {Romero-Shaw}\ \emph {et~al.}(2020)\citenamefont
  {Romero-Shaw}, \citenamefont {Farrow}, \citenamefont {Stevenson},
  \citenamefont {Thrane},\ and\ \citenamefont {Zhu}}]{Romero-Shaw2020a}%
  \BibitemOpen
  \bibfield  {author} {\bibinfo {author} {\bibfnamefont {I.~M.}\ \bibnamefont
  {Romero-Shaw}}, \bibinfo {author} {\bibfnamefont {N.}~\bibnamefont {Farrow}},
  \bibinfo {author} {\bibfnamefont {S.}~\bibnamefont {Stevenson}}, \bibinfo
  {author} {\bibfnamefont {E.}~\bibnamefont {Thrane}}, \ and\ \bibinfo {author}
  {\bibfnamefont {X.~J.}\ \bibnamefont {Zhu}},\ }\href@noop {} {\bibfield
  {journal} {\bibinfo  {journal} {Mon. Not. R. Astron. Soc.}\ }\textbf
  {\bibinfo {volume} {496}},\ \bibinfo {pages} {L64} (\bibinfo {year}
  {2020})}\BibitemShut {NoStop}%
\bibitem [{\citenamefont {Zevin}\ \emph {et~al.}(2020)\citenamefont {Zevin},
  \citenamefont {Berry}, \citenamefont {Coughlin}, \citenamefont
  {Chatziioannou},\ and\ \citenamefont {Vitale}}]{Zevin2020}%
  \BibitemOpen
  \bibfield  {author} {\bibinfo {author} {\bibfnamefont {M.}~\bibnamefont
  {Zevin}}, \bibinfo {author} {\bibfnamefont {C.~P.~L.}\ \bibnamefont {Berry}},
  \bibinfo {author} {\bibfnamefont {S.}~\bibnamefont {Coughlin}}, \bibinfo
  {author} {\bibfnamefont {K.}~\bibnamefont {Chatziioannou}}, \ and\ \bibinfo
  {author} {\bibfnamefont {S.}~\bibnamefont {Vitale}},\ }\href@noop {}
  {\bibfield  {journal} {\bibinfo  {journal} {Astrophys. J.}\ }\textbf
  {\bibinfo {volume} {899}},\ \bibinfo {pages} {L17} (\bibinfo {year}
  {2020})}\BibitemShut {NoStop}%
\bibitem [{\citenamefont {Keitel}(2019)}]{Keitel_2019}%
  \BibitemOpen
  \bibfield  {author} {\bibinfo {author} {\bibfnamefont {D.}~\bibnamefont
  {Keitel}},\ }\href {\doibase 10.3847/2515-5172/ab0c0b} {\bibfield  {journal}
  {\bibinfo  {journal} {Res. Notes. {AAS}}\ }\textbf {\bibinfo {volume} {3}},\
  \bibinfo {pages} {46} (\bibinfo {year} {2019})}\BibitemShut {NoStop}%
\bibitem [{\citenamefont {Ashton}\ and\ \citenamefont
  {Khan}(2020)}]{Ashton2020}%
  \BibitemOpen
  \bibfield  {author} {\bibinfo {author} {\bibfnamefont {G.}~\bibnamefont
  {Ashton}}\ and\ \bibinfo {author} {\bibfnamefont {S.}~\bibnamefont {Khan}},\
  }\href@noop {} {\bibfield  {journal} {\bibinfo  {journal} {Phys. Rev. D}\
  }\textbf {\bibinfo {volume} {101}},\ \bibinfo {pages} {064037} (\bibinfo
  {year} {2020})}\BibitemShut {NoStop}%
\bibitem [{\citenamefont {Payne}\ \emph {et~al.}(2019)\citenamefont {Payne},
  \citenamefont {Talbot},\ and\ \citenamefont {Thrane}}]{Payne2019}%
  \BibitemOpen
  \bibfield  {author} {\bibinfo {author} {\bibfnamefont {E.}~\bibnamefont
  {Payne}}, \bibinfo {author} {\bibfnamefont {C.}~\bibnamefont {Talbot}}, \
  and\ \bibinfo {author} {\bibfnamefont {E.}~\bibnamefont {Thrane}},\
  }\href@noop {} {\bibfield  {journal} {\bibinfo  {journal} {Phys. Rev. D}\
  }\textbf {\bibinfo {volume} {100}},\ \bibinfo {pages} {123017} (\bibinfo
  {year} {2019})}\BibitemShut {NoStop}%
\bibitem [{\citenamefont {Zhao}\ \emph {et~al.}(2019)\citenamefont {Zhao},
  \citenamefont {Lin},\ and\ \citenamefont {Chang}}]{Zhao2019}%
  \BibitemOpen
  \bibfield  {author} {\bibinfo {author} {\bibfnamefont {Z.~C.}\ \bibnamefont
  {Zhao}}, \bibinfo {author} {\bibfnamefont {H.~N.}\ \bibnamefont {Lin}}, \
  and\ \bibinfo {author} {\bibfnamefont {Z.}~\bibnamefont {Chang}},\
  }\href@noop {} {\bibfield  {journal} {\bibinfo  {journal} {Chin. Phys. C}\
  }\textbf {\bibinfo {volume} {43}},\ \bibinfo {pages} {075102} (\bibinfo
  {year} {2019})}\BibitemShut {NoStop}%
\bibitem [{\citenamefont {Wesolowski}\ \emph {et~al.}(2016)\citenamefont
  {Wesolowski}, \citenamefont {Klco}, \citenamefont {Furnstahl}, \citenamefont
  {Phillips},\ and\ \citenamefont {Thapaliya}}]{Wesolowski2016}%
  \BibitemOpen
  \bibfield  {author} {\bibinfo {author} {\bibfnamefont {S.}~\bibnamefont
  {Wesolowski}}, \bibinfo {author} {\bibfnamefont {N.}~\bibnamefont {Klco}},
  \bibinfo {author} {\bibfnamefont {R.~J.}\ \bibnamefont {Furnstahl}}, \bibinfo
  {author} {\bibfnamefont {D.~R.}\ \bibnamefont {Phillips}}, \ and\ \bibinfo
  {author} {\bibfnamefont {A.}~\bibnamefont {Thapaliya}},\ }\href@noop {}
  {\bibfield  {journal} {\bibinfo  {journal} {J. Phys. G}\ }\textbf {\bibinfo
  {volume} {43}},\ \bibinfo {pages} {074001} (\bibinfo {year}
  {2016})}\BibitemShut {NoStop}%
\bibitem [{\citenamefont {Somasundaram}\ \emph {et~al.}(2021)\citenamefont
  {Somasundaram}, \citenamefont {Drischler}, \citenamefont {Tews},\ and\
  \citenamefont {Margueron}}]{Somasundaram2021}%
  \BibitemOpen
  \bibfield  {author} {\bibinfo {author} {\bibfnamefont {R.}~\bibnamefont
  {Somasundaram}}, \bibinfo {author} {\bibfnamefont {C.}~\bibnamefont
  {Drischler}}, \bibinfo {author} {\bibfnamefont {I.}~\bibnamefont {Tews}}, \
  and\ \bibinfo {author} {\bibfnamefont {J.}~\bibnamefont {Margueron}},\
  }\href@noop {} {\bibfield  {journal} {\bibinfo  {journal} {Phys. Rev. C}\
  }\textbf {\bibinfo {volume} {103}},\ \bibinfo {pages} {045803} (\bibinfo
  {year} {2021})}\BibitemShut {NoStop}%
\bibitem [{\citenamefont {Drischler}\ \emph
  {et~al.}(2021{\natexlab{a}})\citenamefont {Drischler}, \citenamefont {Han},
  \citenamefont {Lattimer}, \citenamefont {Prakash}, \citenamefont {Reddy},\
  and\ \citenamefont {Zhao}}]{Drischler2021}%
  \BibitemOpen
  \bibfield  {author} {\bibinfo {author} {\bibfnamefont {C.}~\bibnamefont
  {Drischler}}, \bibinfo {author} {\bibfnamefont {S.}~\bibnamefont {Han}},
  \bibinfo {author} {\bibfnamefont {J.~M.}\ \bibnamefont {Lattimer}}, \bibinfo
  {author} {\bibfnamefont {M.}~\bibnamefont {Prakash}}, \bibinfo {author}
  {\bibfnamefont {S.}~\bibnamefont {Reddy}}, \ and\ \bibinfo {author}
  {\bibfnamefont {T.}~\bibnamefont {Zhao}},\ }\href@noop {} {\bibfield
  {journal} {\bibinfo  {journal} {Phys. Rev. C}\ }\textbf {\bibinfo {volume}
  {103}},\ \bibinfo {pages} {045808} (\bibinfo {year}
  {2021}{\natexlab{a}})}\BibitemShut {NoStop}%
\bibitem [{\citenamefont {Tews}\ \emph {et~al.}(2017)\citenamefont {Tews},
  \citenamefont {Lattimer}, \citenamefont {Ohnishi},\ and\ \citenamefont
  {Kolomeitsev}}]{Tews2017}%
  \BibitemOpen
  \bibfield  {author} {\bibinfo {author} {\bibfnamefont {I.}~\bibnamefont
  {Tews}}, \bibinfo {author} {\bibfnamefont {J.~M.}\ \bibnamefont {Lattimer}},
  \bibinfo {author} {\bibfnamefont {A.}~\bibnamefont {Ohnishi}}, \ and\
  \bibinfo {author} {\bibfnamefont {E.~E.}\ \bibnamefont {Kolomeitsev}},\
  }\href@noop {} {\bibfield  {journal} {\bibinfo  {journal} {Astrophys. J.}\
  }\textbf {\bibinfo {volume} {848}},\ \bibinfo {pages} {105} (\bibinfo {year}
  {2017})}\BibitemShut {NoStop}%
\bibitem [{\citenamefont {Carreau}\ \emph {et~al.}(2019)\citenamefont
  {Carreau}, \citenamefont {Gulminelli},\ and\ \citenamefont
  {Margueron}}]{Carreau2019}%
  \BibitemOpen
  \bibfield  {author} {\bibinfo {author} {\bibfnamefont {T.}~\bibnamefont
  {Carreau}}, \bibinfo {author} {\bibfnamefont {F.}~\bibnamefont {Gulminelli}},
  \ and\ \bibinfo {author} {\bibfnamefont {J.}~\bibnamefont {Margueron}},\
  }\href@noop {} {\bibfield  {journal} {\bibinfo  {journal} {Phys. Rev. C}\
  }\textbf {\bibinfo {volume} {100}},\ \bibinfo {pages} {055803} (\bibinfo
  {year} {2019})}\BibitemShut {NoStop}%
\bibitem [{\citenamefont {Thi}\ \emph {et~al.}(2021{\natexlab{b}})\citenamefont
  {Thi}, \citenamefont {Carreau}, \citenamefont {Fantina},\ and\ \citenamefont
  {Gulminelli}}]{Thi2021a}%
  \BibitemOpen
  \bibfield  {author} {\bibinfo {author} {\bibfnamefont {H.~D.}\ \bibnamefont
  {Thi}}, \bibinfo {author} {\bibfnamefont {T.}~\bibnamefont {Carreau}},
  \bibinfo {author} {\bibfnamefont {A.~F.}\ \bibnamefont {Fantina}}, \ and\
  \bibinfo {author} {\bibfnamefont {F.}~\bibnamefont {Gulminelli}},\
  }\href@noop {} {\bibfield  {journal} {\bibinfo  {journal} {Astron.
  Astrophys.}\ }\textbf {\bibinfo {volume} {654}},\ \bibinfo {pages} {A114}
  (\bibinfo {year} {2021}{\natexlab{b}})}\BibitemShut {NoStop}%
\bibitem [{\citenamefont {Riley}\ \emph {et~al.}(2018)\citenamefont {Riley},
  \citenamefont {Raaijmakers},\ and\ \citenamefont {Watts}}]{Riley2018}%
  \BibitemOpen
  \bibfield  {author} {\bibinfo {author} {\bibfnamefont {T.~E.}\ \bibnamefont
  {Riley}}, \bibinfo {author} {\bibfnamefont {G.}~\bibnamefont {Raaijmakers}},
  \ and\ \bibinfo {author} {\bibfnamefont {A.~L.}\ \bibnamefont {Watts}},\
  }\href@noop {} {\bibfield  {journal} {\bibinfo  {journal} {Mon. Not. R.
  Astron. Soc.}\ }\textbf {\bibinfo {volume} {478}},\ \bibinfo {pages} {1093}
  (\bibinfo {year} {2018})}\BibitemShut {NoStop}%
\bibitem [{\citenamefont {Raaijmakers}\ and\ \citenamefont
  {Others}(2020)}]{Raaijmakers:2019dks}%
  \BibitemOpen
  \bibfield  {author} {\bibinfo {author} {\bibfnamefont {G.}~\bibnamefont
  {Raaijmakers}}\ and\ \bibinfo {author} {\bibnamefont {Others}},\ }\href@noop
  {} {\bibfield  {journal} {\bibinfo  {journal} {Astrophys. J. Lett.}\ }\textbf
  {\bibinfo {volume} {893}},\ \bibinfo {pages} {L21} (\bibinfo {year}
  {2020})}\BibitemShut {NoStop}%
\bibitem [{\citenamefont {Jiang}\ \emph {et~al.}(2020)\citenamefont {Jiang},
  \citenamefont {Tang}, \citenamefont {Wang}, \citenamefont {Fan},\ and\
  \citenamefont {Wei}}]{Jiang:2019rcw}%
  \BibitemOpen
  \bibfield  {author} {\bibinfo {author} {\bibfnamefont {J.-L.}\ \bibnamefont
  {Jiang}}, \bibinfo {author} {\bibfnamefont {S.-P.}\ \bibnamefont {Tang}},
  \bibinfo {author} {\bibfnamefont {Y.-Z.}\ \bibnamefont {Wang}}, \bibinfo
  {author} {\bibfnamefont {Y.-Z.}\ \bibnamefont {Fan}}, \ and\ \bibinfo
  {author} {\bibfnamefont {D.-M.}\ \bibnamefont {Wei}},\ }\href@noop {}
  {\bibfield  {journal} {\bibinfo  {journal} {Astrophys. J.}\ }\textbf
  {\bibinfo {volume} {892}},\ \bibinfo {pages} {1} (\bibinfo {year}
  {2020})}\BibitemShut {NoStop}%
\bibitem [{\citenamefont {G{\"{u}}ven}\ \emph {et~al.}(2020)\citenamefont
  {G{\"{u}}ven}, \citenamefont {Bozkurt}, \citenamefont {Khan},\ and\
  \citenamefont {Margueron}}]{Guven2020}%
  \BibitemOpen
  \bibfield  {author} {\bibinfo {author} {\bibfnamefont {H.}~\bibnamefont
  {G{\"{u}}ven}}, \bibinfo {author} {\bibfnamefont {K.}~\bibnamefont
  {Bozkurt}}, \bibinfo {author} {\bibfnamefont {E.}~\bibnamefont {Khan}}, \
  and\ \bibinfo {author} {\bibfnamefont {J.}~\bibnamefont {Margueron}},\
  }\href@noop {} {\bibfield  {journal} {\bibinfo  {journal} {Phys. Rev. C}\
  }\textbf {\bibinfo {volume} {102}},\ \bibinfo {pages} {015805} (\bibinfo
  {year} {2020})}\BibitemShut {NoStop}%
\bibitem [{\citenamefont {Biswas}(2021)}]{Biswas2021a}%
  \BibitemOpen
  \bibfield  {author} {\bibinfo {author} {\bibfnamefont {B.}~\bibnamefont
  {Biswas}},\ }\href@noop {} {\bibfield  {journal} {\bibinfo  {journal}
  {Astrophys. J.}\ }\textbf {\bibinfo {volume} {921}},\ \bibinfo {pages} {63}
  (\bibinfo {year} {2021})}\BibitemShut {NoStop}%
\bibitem [{\citenamefont {Landry}\ \emph {et~al.}(2020)\citenamefont {Landry},
  \citenamefont {Essick},\ and\ \citenamefont {Chatziioannou}}]{Landry2020a}%
  \BibitemOpen
  \bibfield  {author} {\bibinfo {author} {\bibfnamefont {P.}~\bibnamefont
  {Landry}}, \bibinfo {author} {\bibfnamefont {R.}~\bibnamefont {Essick}}, \
  and\ \bibinfo {author} {\bibfnamefont {K.}~\bibnamefont {Chatziioannou}},\
  }\href@noop {} {\bibfield  {journal} {\bibinfo  {journal} {Phys. Rev. D}\
  }\textbf {\bibinfo {volume} {101}},\ \bibinfo {pages} {123007} (\bibinfo
  {year} {2020})}\BibitemShut {NoStop}%
\bibitem [{\citenamefont {Huth~{\it et. al.}}(2022)}]{Huth2021}%
  \BibitemOpen
  \bibfield  {author} {\bibinfo {author} {\bibfnamefont {S.}~\bibnamefont
  {Huth~{\it et. al.}}},\ }\href {\doibase 10.1038/s41586-022-04750-w}
  {\bibfield  {journal} {\bibinfo  {journal} {Nature}\ }\textbf {\bibinfo
  {volume} {606}},\ \bibinfo {pages} {276} (\bibinfo {year}
  {2022})}\BibitemShut {NoStop}%
\bibitem [{\citenamefont {Biswas}(2022)}]{Biswas2021b}%
  \BibitemOpen
  \bibfield  {author} {\bibinfo {author} {\bibfnamefont {B.}~\bibnamefont
  {Biswas}},\ }\href@noop {} {\bibfield  {journal} {\bibinfo  {journal}
  {Astrophys. J.}\ }\textbf {\bibinfo {volume} {926}},\ \bibinfo {pages} {75}
  (\bibinfo {year} {2022})}\BibitemShut {NoStop}%
\bibitem [{\citenamefont {Imam}\ \emph {et~al.}(2022)\citenamefont {Imam},
  \citenamefont {Patra}, \citenamefont {Mondal}, \citenamefont {Malik},\ and\
  \citenamefont {Agrawal}}]{Imam:2021dbe}%
  \BibitemOpen
  \bibfield  {author} {\bibinfo {author} {\bibfnamefont {S.~M.~A.}\
  \bibnamefont {Imam}}, \bibinfo {author} {\bibfnamefont {N.~K.}\ \bibnamefont
  {Patra}}, \bibinfo {author} {\bibfnamefont {C.}~\bibnamefont {Mondal}},
  \bibinfo {author} {\bibfnamefont {T.}~\bibnamefont {Malik}}, \ and\ \bibinfo
  {author} {\bibfnamefont {B.~K.}\ \bibnamefont {Agrawal}},\ }\href@noop {}
  {\bibfield  {journal} {\bibinfo  {journal} {Phys. Rev. C}\ }\textbf {\bibinfo
  {volume} {105}},\ \bibinfo {pages} {015806} (\bibinfo {year}
  {2022})}\BibitemShut {NoStop}%
\bibitem [{\citenamefont {Chen}\ \emph {et~al.}(2005)\citenamefont {Chen},
  \citenamefont {Ko},\ and\ \citenamefont {Li}}]{Chen:2005ti}%
  \BibitemOpen
  \bibfield  {author} {\bibinfo {author} {\bibfnamefont {L.-W.}\ \bibnamefont
  {Chen}}, \bibinfo {author} {\bibfnamefont {C.~M.}\ \bibnamefont {Ko}}, \ and\
  \bibinfo {author} {\bibfnamefont {B.-A.}\ \bibnamefont {Li}},\ }\href@noop {}
  {\bibfield  {journal} {\bibinfo  {journal} {Phys. Rev. C}\ }\textbf {\bibinfo
  {volume} {72}},\ \bibinfo {pages} {064309} (\bibinfo {year}
  {2005})}\BibitemShut {NoStop}%
\bibitem [{\citenamefont {Chen}\ \emph {et~al.}(2009)\citenamefont {Chen},
  \citenamefont {Cai}, \citenamefont {Ko}, \citenamefont {Li}, \citenamefont
  {Shen},\ and\ \citenamefont {Xu}}]{Chen:2009wv}%
  \BibitemOpen
  \bibfield  {author} {\bibinfo {author} {\bibfnamefont {L.-W.}\ \bibnamefont
  {Chen}}, \bibinfo {author} {\bibfnamefont {B.-J.}\ \bibnamefont {Cai}},
  \bibinfo {author} {\bibfnamefont {C.~M.}\ \bibnamefont {Ko}}, \bibinfo
  {author} {\bibfnamefont {B.-A.}\ \bibnamefont {Li}}, \bibinfo {author}
  {\bibfnamefont {C.}~\bibnamefont {Shen}}, \ and\ \bibinfo {author}
  {\bibfnamefont {J.}~\bibnamefont {Xu}},\ }\href@noop {} {\bibfield  {journal}
  {\bibinfo  {journal} {Phys. Rev. C}\ }\textbf {\bibinfo {volume} {80}},\
  \bibinfo {pages} {014322} (\bibinfo {year} {2009})}\BibitemShut {NoStop}%
\bibitem [{\citenamefont {Newton}\ \emph {et~al.}(2014)\citenamefont {Newton},
  \citenamefont {Hooker}, \citenamefont {Gearheart}, \citenamefont {Murphy},
  \citenamefont {Wen}, \citenamefont {Fattoyev},\ and\ \citenamefont
  {Li}}]{Newton:2014iha}%
  \BibitemOpen
  \bibfield  {author} {\bibinfo {author} {\bibfnamefont {W.~G.}\ \bibnamefont
  {Newton}}, \bibinfo {author} {\bibfnamefont {J.}~\bibnamefont {Hooker}},
  \bibinfo {author} {\bibfnamefont {M.}~\bibnamefont {Gearheart}}, \bibinfo
  {author} {\bibfnamefont {K.}~\bibnamefont {Murphy}}, \bibinfo {author}
  {\bibfnamefont {D.-H.}\ \bibnamefont {Wen}}, \bibinfo {author} {\bibfnamefont
  {F.~J.}\ \bibnamefont {Fattoyev}}, \ and\ \bibinfo {author} {\bibfnamefont
  {B.-A.}\ \bibnamefont {Li}},\ }\href@noop {} {\bibfield  {journal} {\bibinfo
  {journal} {Eur. Phys. J. A}\ }\textbf {\bibinfo {volume} {50}},\ \bibinfo
  {pages} {41} (\bibinfo {year} {2014})}\BibitemShut {NoStop}%
\bibitem [{\citenamefont {Margueron}\ \emph {et~al.}(2018)\citenamefont
  {Margueron}, \citenamefont {{Hoffmann Casali}},\ and\ \citenamefont
  {Gulminelli}}]{Margueron:2017eqc}%
  \BibitemOpen
  \bibfield  {author} {\bibinfo {author} {\bibfnamefont {J.}~\bibnamefont
  {Margueron}}, \bibinfo {author} {\bibfnamefont {R.}~\bibnamefont {{Hoffmann
  Casali}}}, \ and\ \bibinfo {author} {\bibfnamefont {F.}~\bibnamefont
  {Gulminelli}},\ }\href@noop {} {\bibfield  {journal} {\bibinfo  {journal}
  {Phys. Rev. C}\ }\textbf {\bibinfo {volume} {97}},\ \bibinfo {pages} {025805}
  (\bibinfo {year} {2018})}\BibitemShut {NoStop}%
\bibitem [{\citenamefont {Margueron}\ and\ \citenamefont
  {Gulminelli}(2019)}]{Margueron:2018eob}%
  \BibitemOpen
  \bibfield  {author} {\bibinfo {author} {\bibfnamefont {J.}~\bibnamefont
  {Margueron}}\ and\ \bibinfo {author} {\bibfnamefont {F.}~\bibnamefont
  {Gulminelli}},\ }\href@noop {} {\bibfield  {journal} {\bibinfo  {journal}
  {Phys. Rev. C}\ }\textbf {\bibinfo {volume} {99}},\ \bibinfo {pages} {025806}
  (\bibinfo {year} {2019})}\BibitemShut {NoStop}%
\bibitem [{\citenamefont {Lattimer}\ and\ \citenamefont
  {Prakash}(2016)}]{Lattimer2015}%
  \BibitemOpen
  \bibfield  {author} {\bibinfo {author} {\bibfnamefont {J.~M.}\ \bibnamefont
  {Lattimer}}\ and\ \bibinfo {author} {\bibfnamefont {M.}~\bibnamefont
  {Prakash}},\ }\href@noop {} {\bibfield  {journal} {\bibinfo  {journal} {Phys.
  Rept.}\ }\textbf {\bibinfo {volume} {621}},\ \bibinfo {pages} {127} (\bibinfo
  {year} {2016})}\BibitemShut {NoStop}%
\bibitem [{\citenamefont {Gil}\ \emph {et~al.}(2017)\citenamefont {Gil},
  \citenamefont {Papakonstantinou}, \citenamefont {Hyun}, \citenamefont
  {Park},\ and\ \citenamefont {Oh}}]{Gil2017}%
  \BibitemOpen
  \bibfield  {author} {\bibinfo {author} {\bibfnamefont {H.}~\bibnamefont
  {Gil}}, \bibinfo {author} {\bibfnamefont {P.}~\bibnamefont
  {Papakonstantinou}}, \bibinfo {author} {\bibfnamefont {C.~H.}\ \bibnamefont
  {Hyun}}, \bibinfo {author} {\bibfnamefont {T.-S.}\ \bibnamefont {Park}}, \
  and\ \bibinfo {author} {\bibfnamefont {Y.}~\bibnamefont {Oh}},\ }\href@noop
  {} {\bibfield  {journal} {\bibinfo  {journal} {Acta Phys. Polon. B}\ }\textbf
  {\bibinfo {volume} {48}},\ \bibinfo {pages} {305} (\bibinfo {year}
  {2017})}\BibitemShut {NoStop}%
\bibitem [{\citenamefont {Gelman}\ \emph {et~al.}(2013)\citenamefont {Gelman},
  \citenamefont {Carlin}, \citenamefont {Stern}, \citenamefont {Dunson},
  \citenamefont {Vehtari}, \citenamefont {Rubin}, \citenamefont {Carlin},
  \citenamefont {Stern}, \citenamefont {Rubin},\ and\ \citenamefont
  {Dunson}}]{Gelman2013}%
  \BibitemOpen
  \bibfield  {author} {\bibinfo {author} {\bibfnamefont {A.}~\bibnamefont
  {Gelman}}, \bibinfo {author} {\bibfnamefont {J.~B.}\ \bibnamefont {Carlin}},
  \bibinfo {author} {\bibfnamefont {H.~S.}\ \bibnamefont {Stern}}, \bibinfo
  {author} {\bibfnamefont {D.~B.}\ \bibnamefont {Dunson}}, \bibinfo {author}
  {\bibfnamefont {A.}~\bibnamefont {Vehtari}}, \bibinfo {author} {\bibfnamefont
  {D.~B.}\ \bibnamefont {Rubin}}, \bibinfo {author} {\bibfnamefont
  {J.}~\bibnamefont {Carlin}}, \bibinfo {author} {\bibfnamefont
  {H.}~\bibnamefont {Stern}}, \bibinfo {author} {\bibfnamefont
  {D.}~\bibnamefont {Rubin}}, \ and\ \bibinfo {author} {\bibfnamefont
  {D.}~\bibnamefont {Dunson}},\ }\href@noop {} {\emph {\bibinfo {title}
  {{Bayesian Data Analysis Third edition}}}}\ (\bibinfo  {publisher} {CRC
  Press, Boca Raton, Florida},\ \bibinfo {year} {2013})\BibitemShut {NoStop}%
\bibitem [{\citenamefont {Buchner}\ \emph {et~al.}(2014)\citenamefont
  {Buchner}, \citenamefont {Georgakakis}, \citenamefont {Nandra}, \citenamefont
  {Hsu}, \citenamefont {Rangel}, \citenamefont {Brightman}, \citenamefont
  {Merloni}, \citenamefont {Salvato}, \citenamefont {Donley},\ and\
  \citenamefont {Kocevski}}]{Buchner2014}%
  \BibitemOpen
  \bibfield  {author} {\bibinfo {author} {\bibfnamefont {J.}~\bibnamefont
  {Buchner}}, \bibinfo {author} {\bibfnamefont {A.}~\bibnamefont
  {Georgakakis}}, \bibinfo {author} {\bibfnamefont {K.}~\bibnamefont {Nandra}},
  \bibinfo {author} {\bibfnamefont {L.}~\bibnamefont {Hsu}}, \bibinfo {author}
  {\bibfnamefont {C.}~\bibnamefont {Rangel}}, \bibinfo {author} {\bibfnamefont
  {M.}~\bibnamefont {Brightman}}, \bibinfo {author} {\bibfnamefont
  {A.}~\bibnamefont {Merloni}}, \bibinfo {author} {\bibfnamefont
  {M.}~\bibnamefont {Salvato}}, \bibinfo {author} {\bibfnamefont
  {J.}~\bibnamefont {Donley}}, \ and\ \bibinfo {author} {\bibfnamefont
  {D.}~\bibnamefont {Kocevski}},\ }\href {\doibase 10.1051/0004-6361/201322971}
  {\bibfield  {journal} {\bibinfo  {journal} {Astron. Astrophys.}\ }\textbf
  {\bibinfo {volume} {564}},\ \bibinfo {pages} {A125} (\bibinfo {year}
  {2014})}\BibitemShut {NoStop}%
\bibitem [{\citenamefont {Hebeler}\ \emph {et~al.}(2013)\citenamefont
  {Hebeler}, \citenamefont {Lattimer}, \citenamefont {Pethick},\ and\
  \citenamefont {Schwenk}}]{Hebeler:2013nza}%
  \BibitemOpen
  \bibfield  {author} {\bibinfo {author} {\bibfnamefont {K.}~\bibnamefont
  {Hebeler}}, \bibinfo {author} {\bibfnamefont {J.~M.}\ \bibnamefont
  {Lattimer}}, \bibinfo {author} {\bibfnamefont {C.~J.}\ \bibnamefont
  {Pethick}}, \ and\ \bibinfo {author} {\bibfnamefont {A.}~\bibnamefont
  {Schwenk}},\ }\href@noop {} {\bibfield  {journal} {\bibinfo  {journal}
  {Astrophys. J.}\ }\textbf {\bibinfo {volume} {773}},\ \bibinfo {pages} {11}
  (\bibinfo {year} {2013})}\BibitemShut {NoStop}%
\bibitem [{\citenamefont {Chabanat}\ \emph {et~al.}(1998)\citenamefont
  {Chabanat}, \citenamefont {Bonche}, \citenamefont {Haensel}, \citenamefont
  {Meyer},\ and\ \citenamefont {Schaeffer}}]{Chabanat98}%
  \BibitemOpen
  \bibfield  {author} {\bibinfo {author} {\bibfnamefont {E.}~\bibnamefont
  {Chabanat}}, \bibinfo {author} {\bibfnamefont {P.}~\bibnamefont {Bonche}},
  \bibinfo {author} {\bibfnamefont {P.}~\bibnamefont {Haensel}}, \bibinfo
  {author} {\bibfnamefont {J.}~\bibnamefont {Meyer}}, \ and\ \bibinfo {author}
  {\bibfnamefont {R.}~\bibnamefont {Schaeffer}},\ }\href@noop {} {\bibfield
  {journal} {\bibinfo  {journal} {Nucl. Phys. A}\ }\textbf {\bibinfo {volume}
  {635}},\ \bibinfo {pages} {231} (\bibinfo {year} {1998})}\BibitemShut
  {NoStop}%
\bibitem [{\citenamefont {Chabanat}\ \emph {et~al.}(1997)\citenamefont
  {Chabanat}, \citenamefont {Bonche}, \citenamefont {Haensel}, \citenamefont
  {Meyer},\ and\ \citenamefont {Schaeffer}}]{Chabanat97}%
  \BibitemOpen
  \bibfield  {author} {\bibinfo {author} {\bibfnamefont {E.}~\bibnamefont
  {Chabanat}}, \bibinfo {author} {\bibfnamefont {P.}~\bibnamefont {Bonche}},
  \bibinfo {author} {\bibfnamefont {P.}~\bibnamefont {Haensel}}, \bibinfo
  {author} {\bibfnamefont {J.}~\bibnamefont {Meyer}}, \ and\ \bibinfo {author}
  {\bibfnamefont {R.}~\bibnamefont {Schaeffer}},\ }\href@noop {} {\bibfield
  {journal} {\bibinfo  {journal} {Nucl. Phys. A}\ }\textbf {\bibinfo {volume}
  {627}},\ \bibinfo {pages} {710} (\bibinfo {year} {1997})}\BibitemShut
  {NoStop}%
\bibitem [{\citenamefont {Mondal}\ \emph {et~al.}(2015)\citenamefont {Mondal},
  \citenamefont {Agrawal},\ and\ \citenamefont {De}}]{Mondal:2015tfa}%
  \BibitemOpen
  \bibfield  {author} {\bibinfo {author} {\bibfnamefont {C.}~\bibnamefont
  {Mondal}}, \bibinfo {author} {\bibfnamefont {B.~K.}\ \bibnamefont {Agrawal}},
  \ and\ \bibinfo {author} {\bibfnamefont {J.~N.}\ \bibnamefont {De}},\ }\href
  {\doibase 10.1103/PhysRevC.92.024302} {\bibfield  {journal} {\bibinfo
  {journal} {Phys. Rev. C}\ }\textbf {\bibinfo {volume} {92}},\ \bibinfo
  {pages} {024302} (\bibinfo {year} {2015})}\BibitemShut {NoStop}%
\bibitem [{\citenamefont {Mondal}\ \emph {et~al.}(2016)\citenamefont {Mondal},
  \citenamefont {Agrawal}, \citenamefont {De},\ and\ \citenamefont
  {Samaddar}}]{Mondal:2016roo}%
  \BibitemOpen
  \bibfield  {author} {\bibinfo {author} {\bibfnamefont {C.}~\bibnamefont
  {Mondal}}, \bibinfo {author} {\bibfnamefont {B.~K.}\ \bibnamefont {Agrawal}},
  \bibinfo {author} {\bibfnamefont {J.~N.}\ \bibnamefont {De}}, \ and\ \bibinfo
  {author} {\bibfnamefont {S.~K.}\ \bibnamefont {Samaddar}},\ }\href {\doibase
  10.1103/PhysRevC.93.044328} {\bibfield  {journal} {\bibinfo  {journal} {Phys.
  Rev. C}\ }\textbf {\bibinfo {volume} {93}},\ \bibinfo {pages} {044328}
  (\bibinfo {year} {2016})}\BibitemShut {NoStop}%
\bibitem [{\citenamefont {Sulaksono}\ \emph {et~al.}(2009)\citenamefont
  {Sulaksono}, \citenamefont {Buervenich}, \citenamefont {Reinhard},\ and\
  \citenamefont {Maruhn}}]{Sulaksono:2009rn}%
  \BibitemOpen
  \bibfield  {author} {\bibinfo {author} {\bibfnamefont {A.}~\bibnamefont
  {Sulaksono}}, \bibinfo {author} {\bibfnamefont {T.~J.}\ \bibnamefont
  {Buervenich}}, \bibinfo {author} {\bibfnamefont {P.~G.}\ \bibnamefont
  {Reinhard}}, \ and\ \bibinfo {author} {\bibfnamefont {J.~A.}\ \bibnamefont
  {Maruhn}},\ }\href {\doibase 10.1103/PhysRevC.79.044306} {\bibfield
  {journal} {\bibinfo  {journal} {Phys. Rev. C}\ }\textbf {\bibinfo {volume}
  {79}},\ \bibinfo {pages} {044306} (\bibinfo {year} {2009})}\BibitemShut
  {NoStop}%
\bibitem [{\citenamefont {Garg}\ and\ \citenamefont
  {Col\`o}(2018)}]{Garg:2018uam}%
  \BibitemOpen
  \bibfield  {author} {\bibinfo {author} {\bibfnamefont {U.}~\bibnamefont
  {Garg}}\ and\ \bibinfo {author} {\bibfnamefont {G.}~\bibnamefont {Col\`o}},\
  }\href {\doibase 10.1016/j.ppnp.2018.03.001} {\bibfield  {journal} {\bibinfo
  {journal} {Prog. Part. Nucl. Phys.}\ }\textbf {\bibinfo {volume} {101}},\
  \bibinfo {pages} {55} (\bibinfo {year} {2018})}\BibitemShut {NoStop}%
\bibitem [{\citenamefont {Agrawal}\ \emph {et~al.}(2005)\citenamefont
  {Agrawal}, \citenamefont {Shlomo},\ and\ \citenamefont
  {Au}}]{Agrawal:2005ix}%
  \BibitemOpen
  \bibfield  {author} {\bibinfo {author} {\bibfnamefont {B.~K.}\ \bibnamefont
  {Agrawal}}, \bibinfo {author} {\bibfnamefont {S.}~\bibnamefont {Shlomo}}, \
  and\ \bibinfo {author} {\bibfnamefont {V.~K.}\ \bibnamefont {Au}},\ }\href
  {\doibase 10.1103/PhysRevC.72.014310} {\bibfield  {journal} {\bibinfo
  {journal} {Phys. Rev. C}\ }\textbf {\bibinfo {volume} {72}},\ \bibinfo
  {pages} {014310} (\bibinfo {year} {2005})}\BibitemShut {NoStop}%
\bibitem [{\citenamefont {Reed}\ \emph {et~al.}(2021)\citenamefont {Reed},
  \citenamefont {Fattoyev}, \citenamefont {Horowitz},\ and\ \citenamefont
  {Piekarewicz}}]{Reed:2021nqk}%
  \BibitemOpen
  \bibfield  {author} {\bibinfo {author} {\bibfnamefont {B.~T.}\ \bibnamefont
  {Reed}}, \bibinfo {author} {\bibfnamefont {F.~J.}\ \bibnamefont {Fattoyev}},
  \bibinfo {author} {\bibfnamefont {C.~J.}\ \bibnamefont {Horowitz}}, \ and\
  \bibinfo {author} {\bibfnamefont {J.}~\bibnamefont {Piekarewicz}},\
  }\href@noop {} {\bibfield  {journal} {\bibinfo  {journal} {Phys. Rev. Lett.}\
  }\textbf {\bibinfo {volume} {126}},\ \bibinfo {pages} {172503} (\bibinfo
  {year} {2021})}\BibitemShut {NoStop}%
\bibitem [{\citenamefont {Roca-Maza}\ \emph {et~al.}(2015)\citenamefont
  {Roca-Maza}, \citenamefont {Vi\~nas}, \citenamefont {Centelles},
  \citenamefont {Agrawal}, \citenamefont {Colo'}, \citenamefont {Paar},
  \citenamefont {Piekarewicz},\ and\ \citenamefont
  {Vretenar}}]{Roca-Maza:2015eza}%
  \BibitemOpen
  \bibfield  {author} {\bibinfo {author} {\bibfnamefont {X.}~\bibnamefont
  {Roca-Maza}}, \bibinfo {author} {\bibfnamefont {X.}~\bibnamefont {Vi\~nas}},
  \bibinfo {author} {\bibfnamefont {M.}~\bibnamefont {Centelles}}, \bibinfo
  {author} {\bibfnamefont {B.~K.}\ \bibnamefont {Agrawal}}, \bibinfo {author}
  {\bibfnamefont {G.}~\bibnamefont {Colo'}}, \bibinfo {author} {\bibfnamefont
  {N.}~\bibnamefont {Paar}}, \bibinfo {author} {\bibfnamefont {J.}~\bibnamefont
  {Piekarewicz}}, \ and\ \bibinfo {author} {\bibfnamefont {D.}~\bibnamefont
  {Vretenar}},\ }\href {\doibase 10.1103/PhysRevC.92.064304} {\bibfield
  {journal} {\bibinfo  {journal} {Phys. Rev. C}\ }\textbf {\bibinfo {volume}
  {92}},\ \bibinfo {pages} {064304} (\bibinfo {year} {2015})}\BibitemShut
  {NoStop}%
\bibitem [{\citenamefont {Essick}\ \emph {et~al.}(2021)\citenamefont {Essick},
  \citenamefont {Tews}, \citenamefont {Landry},\ and\ \citenamefont
  {Schwenk}}]{Essick:2021kjb}%
  \BibitemOpen
  \bibfield  {author} {\bibinfo {author} {\bibfnamefont {R.}~\bibnamefont
  {Essick}}, \bibinfo {author} {\bibfnamefont {I.}~\bibnamefont {Tews}},
  \bibinfo {author} {\bibfnamefont {P.}~\bibnamefont {Landry}}, \ and\ \bibinfo
  {author} {\bibfnamefont {A.}~\bibnamefont {Schwenk}},\ }\href {\doibase
  10.1103/PhysRevLett.127.192701} {\bibfield  {journal} {\bibinfo  {journal}
  {Phys. Rev. Lett.}\ }\textbf {\bibinfo {volume} {127}},\ \bibinfo {pages}
  {192701} (\bibinfo {year} {2021})}\BibitemShut {NoStop}%
\bibitem [{\citenamefont {Ferreira}\ and\ \citenamefont
  {Provid\^encia}(2021)}]{Ferreira:2021pni}%
  \BibitemOpen
  \bibfield  {author} {\bibinfo {author} {\bibfnamefont {M.}~\bibnamefont
  {Ferreira}}\ and\ \bibinfo {author} {\bibfnamefont {C.}~\bibnamefont
  {Provid\^encia}},\ }\href {\doibase 10.1103/PhysRevD.104.063006} {\bibfield
  {journal} {\bibinfo  {journal} {Phys. Rev. D}\ }\textbf {\bibinfo {volume}
  {104}},\ \bibinfo {pages} {063006} (\bibinfo {year} {2021})}\BibitemShut
  {NoStop}%
\bibitem [{\citenamefont {Dutra}\ \emph {et~al.}(2012)\citenamefont {Dutra},
  \citenamefont {Lourenco}, \citenamefont {Sa~Martins}, \citenamefont
  {Delfino}, \citenamefont {Stone},\ and\ \citenamefont
  {Stevenson}}]{Dutra:2012mb}%
  \BibitemOpen
  \bibfield  {author} {\bibinfo {author} {\bibfnamefont {M.}~\bibnamefont
  {Dutra}}, \bibinfo {author} {\bibfnamefont {O.}~\bibnamefont {Lourenco}},
  \bibinfo {author} {\bibfnamefont {J.~S.}\ \bibnamefont {Sa~Martins}},
  \bibinfo {author} {\bibfnamefont {A.}~\bibnamefont {Delfino}}, \bibinfo
  {author} {\bibfnamefont {J.~R.}\ \bibnamefont {Stone}}, \ and\ \bibinfo
  {author} {\bibfnamefont {P.~D.}\ \bibnamefont {Stevenson}},\ }\href {\doibase
  10.1103/PhysRevC.85.035201} {\bibfield  {journal} {\bibinfo  {journal} {Phys.
  Rev. C}\ }\textbf {\bibinfo {volume} {85}},\ \bibinfo {pages} {035201}
  (\bibinfo {year} {2012})}\BibitemShut {NoStop}%
\bibitem [{\citenamefont {Dutra}\ \emph {et~al.}(2014)\citenamefont {Dutra},
  \citenamefont {Louren\c{c}o}, \citenamefont {Avancini}, \citenamefont
  {Carlson}, \citenamefont {Delfino}, \citenamefont {Menezes}, \citenamefont
  {Provid\^encia}, \citenamefont {Typel},\ and\ \citenamefont
  {Stone}}]{Dutra:2014qga}%
  \BibitemOpen
  \bibfield  {author} {\bibinfo {author} {\bibfnamefont {M.}~\bibnamefont
  {Dutra}}, \bibinfo {author} {\bibfnamefont {O.}~\bibnamefont {Louren\c{c}o}},
  \bibinfo {author} {\bibfnamefont {S.~S.}\ \bibnamefont {Avancini}}, \bibinfo
  {author} {\bibfnamefont {B.~V.}\ \bibnamefont {Carlson}}, \bibinfo {author}
  {\bibfnamefont {A.}~\bibnamefont {Delfino}}, \bibinfo {author} {\bibfnamefont
  {D.~P.}\ \bibnamefont {Menezes}}, \bibinfo {author} {\bibfnamefont
  {C.}~\bibnamefont {Provid\^encia}}, \bibinfo {author} {\bibfnamefont
  {S.}~\bibnamefont {Typel}}, \ and\ \bibinfo {author} {\bibfnamefont {J.~R.}\
  \bibnamefont {Stone}},\ }\href {\doibase 10.1103/PhysRevC.90.055203}
  {\bibfield  {journal} {\bibinfo  {journal} {Phys. Rev. C}\ }\textbf {\bibinfo
  {volume} {90}},\ \bibinfo {pages} {055203} (\bibinfo {year}
  {2014})}\BibitemShut {NoStop}%
\bibitem [{\citenamefont {Mondal}\ \emph {et~al.}(2017)\citenamefont {Mondal},
  \citenamefont {Agrawal}, \citenamefont {De}, \citenamefont {Samaddar},
  \citenamefont {Centelles},\ and\ \citenamefont
  {Vi{\~{n}}as}}]{Mondal:2017hnh}%
  \BibitemOpen
  \bibfield  {author} {\bibinfo {author} {\bibfnamefont {C.}~\bibnamefont
  {Mondal}}, \bibinfo {author} {\bibfnamefont {B.~K.}\ \bibnamefont {Agrawal}},
  \bibinfo {author} {\bibfnamefont {J.~N.}\ \bibnamefont {De}}, \bibinfo
  {author} {\bibfnamefont {S.~K.}\ \bibnamefont {Samaddar}}, \bibinfo {author}
  {\bibfnamefont {M.}~\bibnamefont {Centelles}}, \ and\ \bibinfo {author}
  {\bibfnamefont {X.}~\bibnamefont {Vi{\~{n}}as}},\ }\href@noop {} {\bibfield
  {journal} {\bibinfo  {journal} {Phys. Rev. C}\ }\textbf {\bibinfo {volume}
  {96}},\ \bibinfo {pages} {021302} (\bibinfo {year} {2017})}\BibitemShut
  {NoStop}%
\bibitem [{\citenamefont {Drischler}\ \emph
  {et~al.}(2021{\natexlab{b}})\citenamefont {Drischler}, \citenamefont {Holt},\
  and\ \citenamefont {Wellenhofer}}]{Drischler:2021kxf}%
  \BibitemOpen
  \bibfield  {author} {\bibinfo {author} {\bibfnamefont {C.}~\bibnamefont
  {Drischler}}, \bibinfo {author} {\bibfnamefont {J.~W.}\ \bibnamefont {Holt}},
  \ and\ \bibinfo {author} {\bibfnamefont {C.}~\bibnamefont {Wellenhofer}},\
  }\href {\doibase 10.1146/annurev-nucl-102419-041903} {\bibfield  {journal}
  {\bibinfo  {journal} {Ann. Rev. Nucl. Part. Sci.}\ }\textbf {\bibinfo
  {volume} {71}},\ \bibinfo {pages} {403} (\bibinfo {year}
  {2021}{\natexlab{b}})}\BibitemShut {NoStop}%
\bibitem [{\citenamefont {Ekstr\"om}\ \emph {et~al.}(2015)\citenamefont
  {Ekstr\"om}, \citenamefont {Jansen}, \citenamefont {Wendt}, \citenamefont
  {Hagen}, \citenamefont {Papenbrock}, \citenamefont {Carlsson}, \citenamefont
  {Forss\'en}, \citenamefont {Hjorth-Jensen}, \citenamefont {Navr\'atil},\ and\
  \citenamefont {Nazarewicz}}]{Ekstrom:2015rta}%
  \BibitemOpen
  \bibfield  {author} {\bibinfo {author} {\bibfnamefont {A.}~\bibnamefont
  {Ekstr\"om}}, \bibinfo {author} {\bibfnamefont {G.~R.}\ \bibnamefont
  {Jansen}}, \bibinfo {author} {\bibfnamefont {K.~A.}\ \bibnamefont {Wendt}},
  \bibinfo {author} {\bibfnamefont {G.}~\bibnamefont {Hagen}}, \bibinfo
  {author} {\bibfnamefont {T.}~\bibnamefont {Papenbrock}}, \bibinfo {author}
  {\bibfnamefont {B.~D.}\ \bibnamefont {Carlsson}}, \bibinfo {author}
  {\bibfnamefont {C.}~\bibnamefont {Forss\'en}}, \bibinfo {author}
  {\bibfnamefont {M.}~\bibnamefont {Hjorth-Jensen}}, \bibinfo {author}
  {\bibfnamefont {P.}~\bibnamefont {Navr\'atil}}, \ and\ \bibinfo {author}
  {\bibfnamefont {W.}~\bibnamefont {Nazarewicz}},\ }\href {\doibase
  10.1103/PhysRevC.91.051301} {\bibfield  {journal} {\bibinfo  {journal} {Phys.
  Rev. C}\ }\textbf {\bibinfo {volume} {91}},\ \bibinfo {pages} {051301}
  (\bibinfo {year} {2015})}\BibitemShut {NoStop}%
\bibitem [{\citenamefont {Lim}\ and\ \citenamefont {Holt}(2019)}]{Lim:2019som}%
  \BibitemOpen
  \bibfield  {author} {\bibinfo {author} {\bibfnamefont {Y.}~\bibnamefont
  {Lim}}\ and\ \bibinfo {author} {\bibfnamefont {J.~W.}\ \bibnamefont {Holt}},\
  }\href {\doibase 10.1140/epja/i2019-12917-9} {\bibfield  {journal} {\bibinfo
  {journal} {Eur. Phys. J. A}\ }\textbf {\bibinfo {volume} {55}},\ \bibinfo
  {pages} {209} (\bibinfo {year} {2019})}\BibitemShut {NoStop}%
\bibitem [{\citenamefont {Malik}\ \emph {et~al.}(2022)\citenamefont {Malik},
  \citenamefont {Ferreira}, \citenamefont {Agrawal},\ and\ \citenamefont
  {Provid\^encia}}]{Malik:2022zol}%
  \BibitemOpen
  \bibfield  {author} {\bibinfo {author} {\bibfnamefont {T.}~\bibnamefont
  {Malik}}, \bibinfo {author} {\bibfnamefont {M.}~\bibnamefont {Ferreira}},
  \bibinfo {author} {\bibfnamefont {B.~K.}\ \bibnamefont {Agrawal}}, \ and\
  \bibinfo {author} {\bibfnamefont {C.}~\bibnamefont {Provid\^encia}},\ }\href
  {\doibase 10.3847/1538-4357/ac5d3c} {\bibfield  {journal} {\bibinfo
  {journal} {Astrophys. J.}\ }\textbf {\bibinfo {volume} {930}},\ \bibinfo
  {pages} {17} (\bibinfo {year} {2022})}\BibitemShut {NoStop}%
\bibitem [{\citenamefont {Ghosh}\ \emph {et~al.}(2022)\citenamefont {Ghosh},
  \citenamefont {Pradhan}, \citenamefont {Chatterjee},\ and\ \citenamefont
  {Schaffner-Bielich}}]{Ghosh:2022lam}%
  \BibitemOpen
  \bibfield  {author} {\bibinfo {author} {\bibfnamefont {S.}~\bibnamefont
  {Ghosh}}, \bibinfo {author} {\bibfnamefont {B.~K.}\ \bibnamefont {Pradhan}},
  \bibinfo {author} {\bibfnamefont {D.}~\bibnamefont {Chatterjee}}, \ and\
  \bibinfo {author} {\bibfnamefont {J.}~\bibnamefont {Schaffner-Bielich}},\
  }\href {\doibase 10.3389/fspas.2022.864294} {\bibfield  {journal} {\bibinfo
  {journal} {Front. Astron. Space Sci.}\ }\textbf {\bibinfo {volume} {9}},\
  \bibinfo {pages} {864294} (\bibinfo {year} {2022})}\BibitemShut {NoStop}%
\bibitem [{\citenamefont {Glendenning}(1992)}]{Glendenning:1992dr}%
  \BibitemOpen
  \bibfield  {author} {\bibinfo {author} {\bibfnamefont {N.~K.}\ \bibnamefont
  {Glendenning}},\ }\href {\doibase 10.1103/PhysRevD.46.4161} {\bibfield
  {journal} {\bibinfo  {journal} {Phys. Rev. D}\ }\textbf {\bibinfo {volume}
  {46}},\ \bibinfo {pages} {4161} (\bibinfo {year} {1992})}\BibitemShut
  {NoStop}%
\bibitem [{\citenamefont {Adhikari~{\sl et al.}}(2021)}]{PREX:2021umo}%
  \BibitemOpen
  \bibfield  {author} {\bibinfo {author} {\bibfnamefont {D.}~\bibnamefont
  {Adhikari~{\sl et al.}}} (\bibinfo {collaboration} {PREX}),\ }\href@noop {}
  {\bibfield  {journal} {\bibinfo  {journal} {Phys. Rev. Lett.}\ }\textbf
  {\bibinfo {volume} {126}},\ \bibinfo {pages} {172502} (\bibinfo {year}
  {2021})}\BibitemShut {NoStop}%
\bibitem [{\citenamefont {Oppenheimer}\ and\ \citenamefont
  {Volkoff}(1939)}]{Oppenheimer:1939ne}%
  \BibitemOpen
  \bibfield  {author} {\bibinfo {author} {\bibfnamefont {J.~R.}\ \bibnamefont
  {Oppenheimer}}\ and\ \bibinfo {author} {\bibfnamefont {G.~M.}\ \bibnamefont
  {Volkoff}},\ }\href@noop {} {\bibfield  {journal} {\bibinfo  {journal} {Phys.
  Rev.}\ }\textbf {\bibinfo {volume} {55}},\ \bibinfo {pages} {374} (\bibinfo
  {year} {1939})}\BibitemShut {NoStop}%
\bibitem [{\citenamefont {Tolman}(1939)}]{Tolman:1939jz}%
  \BibitemOpen
  \bibfield  {author} {\bibinfo {author} {\bibfnamefont {R.~C.}\ \bibnamefont
  {Tolman}},\ }\href@noop {} {\bibfield  {journal} {\bibinfo  {journal} {Phys.
  Rev.}\ }\textbf {\bibinfo {volume} {55}},\ \bibinfo {pages} {364} (\bibinfo
  {year} {1939})}\BibitemShut {NoStop}%
\bibitem [{\citenamefont {Baym}\ \emph {et~al.}(1971)\citenamefont {Baym},
  \citenamefont {Pethick},\ and\ \citenamefont {Sutherland}}]{Baym:1971pw}%
  \BibitemOpen
  \bibfield  {author} {\bibinfo {author} {\bibfnamefont {G.}~\bibnamefont
  {Baym}}, \bibinfo {author} {\bibfnamefont {C.}~\bibnamefont {Pethick}}, \
  and\ \bibinfo {author} {\bibfnamefont {P.}~\bibnamefont {Sutherland}},\
  }\href@noop {} {\bibfield  {journal} {\bibinfo  {journal} {Astrophys. J.}\
  }\textbf {\bibinfo {volume} {170}},\ \bibinfo {pages} {299} (\bibinfo {year}
  {1971})}\BibitemShut {NoStop}%
\bibitem [{\citenamefont {Carriere}\ \emph {et~al.}(2003)\citenamefont
  {Carriere}, \citenamefont {Horowitz},\ and\ \citenamefont
  {Piekarewicz}}]{Carriere:2002bx}%
  \BibitemOpen
  \bibfield  {author} {\bibinfo {author} {\bibfnamefont {J.}~\bibnamefont
  {Carriere}}, \bibinfo {author} {\bibfnamefont {C.~J.}\ \bibnamefont
  {Horowitz}}, \ and\ \bibinfo {author} {\bibfnamefont {J.}~\bibnamefont
  {Piekarewicz}},\ }\href@noop {} {\bibfield  {journal} {\bibinfo  {journal}
  {Astrophys. J.}\ }\textbf {\bibinfo {volume} {593}},\ \bibinfo {pages} {463}
  (\bibinfo {year} {2003})}\BibitemShut {NoStop}%
\bibitem [{\citenamefont {Fortin}\ \emph {et~al.}(2016)\citenamefont {Fortin},
  \citenamefont {Providencia}, \citenamefont {Raduta}, \citenamefont
  {Gulminelli}, \citenamefont {Zdunik}, \citenamefont {Haensel},\ and\
  \citenamefont {Bejger}}]{Fortin:2016hny}%
  \BibitemOpen
  \bibfield  {author} {\bibinfo {author} {\bibfnamefont {M.}~\bibnamefont
  {Fortin}}, \bibinfo {author} {\bibfnamefont {C.}~\bibnamefont {Providencia}},
  \bibinfo {author} {\bibfnamefont {A.~R.}\ \bibnamefont {Raduta}}, \bibinfo
  {author} {\bibfnamefont {F.}~\bibnamefont {Gulminelli}}, \bibinfo {author}
  {\bibfnamefont {J.~L.}\ \bibnamefont {Zdunik}}, \bibinfo {author}
  {\bibfnamefont {P.}~\bibnamefont {Haensel}}, \ and\ \bibinfo {author}
  {\bibfnamefont {M.}~\bibnamefont {Bejger}},\ }\href@noop {} {\bibfield
  {journal} {\bibinfo  {journal} {Phys. Rev. C}\ }\textbf {\bibinfo {volume}
  {94}},\ \bibinfo {pages} {035804} (\bibinfo {year} {2016})}\BibitemShut
  {NoStop}%
\bibitem [{\citenamefont {Piekarewicz}\ and\ \citenamefont
  {Fattoyev}(2019{\natexlab{b}})}]{Piekarewicz:2018sgy}%
  \BibitemOpen
  \bibfield  {author} {\bibinfo {author} {\bibfnamefont {J.}~\bibnamefont
  {Piekarewicz}}\ and\ \bibinfo {author} {\bibfnamefont {F.~J.}\ \bibnamefont
  {Fattoyev}},\ }\href@noop {} {\bibfield  {journal} {\bibinfo  {journal}
  {Phys. Rev. C}\ }\textbf {\bibinfo {volume} {99}},\ \bibinfo {pages} {045802}
  (\bibinfo {year} {2019}{\natexlab{b}})}\BibitemShut {NoStop}%
\bibitem [{\citenamefont {Abbott~{\sl et
  al.}}(2018{\natexlab{b}})}]{GW170817_MR_PEsample}%
  \BibitemOpen
  \bibfield  {author} {\bibinfo {author} {\bibfnamefont {B.~P.}\ \bibnamefont
  {Abbott~{\sl et al.}}},\ }\href@noop {} {\bibfield  {journal} {\bibinfo
  {journal} {Phys. Rev. Lett.}\ }\textbf {\bibinfo {volume} {121}},\ \bibinfo
  {pages} {161101} (\bibinfo {year} {2018}{\natexlab{b}})}\BibitemShut
  {NoStop}%
\bibitem [{\citenamefont {Abbott~{\sl et
  al.}}(2019{\natexlab{c}})}]{Abbott2019a}%
  \BibitemOpen
  \bibfield  {author} {\bibinfo {author} {\bibfnamefont {B.~P.}\ \bibnamefont
  {Abbott~{\sl et al.}}},\ }\href@noop {} {\bibfield  {journal} {\bibinfo
  {journal} {Phys. Rev. X}\ }\textbf {\bibinfo {volume} {9}},\ \bibinfo {pages}
  {011001} (\bibinfo {year} {2019}{\natexlab{c}})}\BibitemShut {NoStop}%
\bibitem [{\citenamefont {{Abbott}~{\sl et al.}}(2021)}]{GWOSC_softx}%
  \BibitemOpen
  \bibfield  {author} {\bibinfo {author} {\bibfnamefont {R.}~\bibnamefont
  {{Abbott}~{\sl et al.}}},\ }\href@noop {} {\bibfield  {journal} {\bibinfo
  {journal} {SoftwareX}\ }\textbf {\bibinfo {volume} {13}},\ \bibinfo {eid}
  {100658} (\bibinfo {year} {2021})}\BibitemShut {NoStop}%
\bibitem [{\citenamefont {Riley~{\sl et
  al.}}(2019{\natexlab{b}})}]{Riley:2019yda}%
  \BibitemOpen
  \bibfield  {author} {\bibinfo {author} {\bibfnamefont {T.~E.}\ \bibnamefont
  {Riley~{\sl et al.}}},\ }\href@noop {} {\bibfield  {journal} {\bibinfo
  {journal} {Astrophys. J. Lett.}\ }\textbf {\bibinfo {volume} {887}},\
  \bibinfo {pages} {L21} (\bibinfo {year} {2019}{\natexlab{b}})}\BibitemShut
  {NoStop}%
\bibitem [{\citenamefont {Miller~{\sl et
  al.}}(2019{\natexlab{b}})}]{Miller:2019cac}%
  \BibitemOpen
  \bibfield  {author} {\bibinfo {author} {\bibfnamefont {M.~C.}\ \bibnamefont
  {Miller~{\sl et al.}}},\ }\href@noop {} {\bibfield  {journal} {\bibinfo
  {journal} {Astrophys. J. Lett.}\ }\textbf {\bibinfo {volume} {887}},\
  \bibinfo {pages} {L24} (\bibinfo {year} {2019}{\natexlab{b}})}\BibitemShut
  {NoStop}%
\bibitem [{\citenamefont {Riley~{\sl et al.}}(2021)}]{Riley:2021pdl}%
  \BibitemOpen
  \bibfield  {author} {\bibinfo {author} {\bibfnamefont {T.~E.}\ \bibnamefont
  {Riley~{\sl et al.}}},\ }\href@noop {} {\bibfield  {journal} {\bibinfo
  {journal} {Astrophys. J. Lett.}\ }\textbf {\bibinfo {volume} {918}},\
  \bibinfo {pages} {L27} (\bibinfo {year} {2021})}\BibitemShut {NoStop}%
\bibitem [{\citenamefont {Miller~{\sl et al.}}(2021)}]{Miller:2021qha}%
  \BibitemOpen
  \bibfield  {author} {\bibinfo {author} {\bibfnamefont {M.~C.}\ \bibnamefont
  {Miller~{\sl et al.}}},\ }\href@noop {} {\bibfield  {journal} {\bibinfo
  {journal} {Astrophys. J. Lett.}\ }\textbf {\bibinfo {volume} {918}},\
  \bibinfo {pages} {L28} (\bibinfo {year} {2021})}\BibitemShut {NoStop}%
\bibitem [{\citenamefont {Romani}\ \emph {et~al.}(2021)\citenamefont {Romani},
  \citenamefont {Kandel}, \citenamefont {Filippenko}, \citenamefont {Brink},\
  and\ \citenamefont {Zheng}}]{Romani:2021xmb}%
  \BibitemOpen
  \bibfield  {author} {\bibinfo {author} {\bibfnamefont {R.~W.}\ \bibnamefont
  {Romani}}, \bibinfo {author} {\bibfnamefont {D.}~\bibnamefont {Kandel}},
  \bibinfo {author} {\bibfnamefont {A.~V.}\ \bibnamefont {Filippenko}},
  \bibinfo {author} {\bibfnamefont {T.~G.}\ \bibnamefont {Brink}}, \ and\
  \bibinfo {author} {\bibfnamefont {W.}~\bibnamefont {Zheng}},\ }\href@noop {}
  {\bibfield  {journal} {\bibinfo  {journal} {Astrophys. J. Lett.}\ }\textbf
  {\bibinfo {volume} {908}},\ \bibinfo {pages} {L46} (\bibinfo {year}
  {2021})}\BibitemShut {NoStop}%
\bibitem [{\citenamefont {Rezzolla}\ \emph {et~al.}(2018)\citenamefont
  {Rezzolla}, \citenamefont {Most},\ and\ \citenamefont
  {Weih}}]{Rezzolla_2018}%
  \BibitemOpen
  \bibfield  {author} {\bibinfo {author} {\bibfnamefont {L.}~\bibnamefont
  {Rezzolla}}, \bibinfo {author} {\bibfnamefont {E.~R.}\ \bibnamefont {Most}},
  \ and\ \bibinfo {author} {\bibfnamefont {L.~R.}\ \bibnamefont {Weih}},\
  }\href {\doibase 10.3847/2041-8213/aaa401} {\bibfield  {journal} {\bibinfo
  {journal} {Astrophys. J. Lett.}\ }\textbf {\bibinfo {volume} {852}},\
  \bibinfo {pages} {L25} (\bibinfo {year} {2018})}\BibitemShut {NoStop}%
\bibitem [{\citenamefont {Abbott~{\sl et al.}}(2020{\natexlab{b}})}]{GW190814}%
  \BibitemOpen
  \bibfield  {author} {\bibinfo {author} {\bibfnamefont {R.}~\bibnamefont
  {Abbott~{\sl et al.}}} (\bibinfo {collaboration} {LIGO Scientific, Virgo}),\
  }\href {\doibase 10.3847/2041-8213/ab960f} {\bibfield  {journal} {\bibinfo
  {journal} {Astrophys. J. Lett.}\ }\textbf {\bibinfo {volume} {896}},\
  \bibinfo {pages} {L44} (\bibinfo {year} {2020}{\natexlab{b}})}\BibitemShut
  {NoStop}%
\bibitem [{\citenamefont {Tsokaros}\ \emph {et~al.}(2020)\citenamefont
  {Tsokaros}, \citenamefont {Ruiz},\ and\ \citenamefont
  {Shapiro}}]{Tsokaros_2020}%
  \BibitemOpen
  \bibfield  {author} {\bibinfo {author} {\bibfnamefont {A.}~\bibnamefont
  {Tsokaros}}, \bibinfo {author} {\bibfnamefont {M.}~\bibnamefont {Ruiz}}, \
  and\ \bibinfo {author} {\bibfnamefont {S.~L.}\ \bibnamefont {Shapiro}},\
  }\href {\doibase 10.3847/1538-4357/abc421} {\bibfield  {journal} {\bibinfo
  {journal} {Astrophys. J.}\ }\textbf {\bibinfo {volume} {905}},\ \bibinfo
  {pages} {48} (\bibinfo {year} {2020})}\BibitemShut {NoStop}%
\bibitem [{\citenamefont {Lim}\ \emph {et~al.}(2021)\citenamefont {Lim},
  \citenamefont {Bhattacharya}, \citenamefont {Holt},\ and\ \citenamefont
  {Pati}}]{Lim_2021}%
  \BibitemOpen
  \bibfield  {author} {\bibinfo {author} {\bibfnamefont {Y.}~\bibnamefont
  {Lim}}, \bibinfo {author} {\bibfnamefont {A.}~\bibnamefont {Bhattacharya}},
  \bibinfo {author} {\bibfnamefont {J.~W.}\ \bibnamefont {Holt}}, \ and\
  \bibinfo {author} {\bibfnamefont {D.}~\bibnamefont {Pati}},\ }\href {\doibase
  10.1103/PhysRevC.104.L032802} {\bibfield  {journal} {\bibinfo  {journal}
  {Phys. Rev. C}\ }\textbf {\bibinfo {volume} {104}},\ \bibinfo {pages}
  {L032802} (\bibinfo {year} {2021})}\BibitemShut {NoStop}%
\bibitem [{\citenamefont {Drischler}\ \emph
  {et~al.}(2021{\natexlab{c}})\citenamefont {Drischler}, \citenamefont {Han},
  \citenamefont {Lattimer}, \citenamefont {Prakash}, \citenamefont {Reddy},\
  and\ \citenamefont {Zhao}}]{Drischler_2021}%
  \BibitemOpen
  \bibfield  {author} {\bibinfo {author} {\bibfnamefont {C.}~\bibnamefont
  {Drischler}}, \bibinfo {author} {\bibfnamefont {S.}~\bibnamefont {Han}},
  \bibinfo {author} {\bibfnamefont {J.~M.}\ \bibnamefont {Lattimer}}, \bibinfo
  {author} {\bibfnamefont {M.}~\bibnamefont {Prakash}}, \bibinfo {author}
  {\bibfnamefont {S.}~\bibnamefont {Reddy}}, \ and\ \bibinfo {author}
  {\bibfnamefont {T.}~\bibnamefont {Zhao}},\ }\href {\doibase
  10.1103/PhysRevC.103.045808} {\bibfield  {journal} {\bibinfo  {journal}
  {Phys. Rev. C}\ }\textbf {\bibinfo {volume} {103}},\ \bibinfo {pages}
  {045808} (\bibinfo {year} {2021}{\natexlab{c}})}\BibitemShut {NoStop}%
\bibitem [{\citenamefont {Li}\ \emph {et~al.}(2021)\citenamefont {Li},
  \citenamefont {Miao}, \citenamefont {Han},\ and\ \citenamefont
  {Zhang}}]{Li:2021crp}%
  \BibitemOpen
  \bibfield  {author} {\bibinfo {author} {\bibfnamefont {A.}~\bibnamefont
  {Li}}, \bibinfo {author} {\bibfnamefont {Z.}~\bibnamefont {Miao}}, \bibinfo
  {author} {\bibfnamefont {S.}~\bibnamefont {Han}}, \ and\ \bibinfo {author}
  {\bibfnamefont {B.}~\bibnamefont {Zhang}},\ }\href {\doibase
  10.3847/1538-4357/abf355} {\bibfield  {journal} {\bibinfo  {journal}
  {Astrophys. J.}\ }\textbf {\bibinfo {volume} {913}},\ \bibinfo {pages} {27}
  (\bibinfo {year} {2021})}\BibitemShut {NoStop}%
\bibitem [{\citenamefont {Ferreira}\ \emph {et~al.}(2020)\citenamefont
  {Ferreira}, \citenamefont {Fortin}, \citenamefont {Malik}, \citenamefont
  {Agrawal},\ and\ \citenamefont {Provid\^encia}}]{Ferreira:2019bgy}%
  \BibitemOpen
  \bibfield  {author} {\bibinfo {author} {\bibfnamefont {M.}~\bibnamefont
  {Ferreira}}, \bibinfo {author} {\bibfnamefont {M.}~\bibnamefont {Fortin}},
  \bibinfo {author} {\bibfnamefont {T.}~\bibnamefont {Malik}}, \bibinfo
  {author} {\bibfnamefont {B.~K.}\ \bibnamefont {Agrawal}}, \ and\ \bibinfo
  {author} {\bibfnamefont {C.}~\bibnamefont {Provid\^encia}},\ }\href@noop {}
  {\bibfield  {journal} {\bibinfo  {journal} {Phys. Rev. D}\ }\textbf {\bibinfo
  {volume} {101}},\ \bibinfo {pages} {043021} (\bibinfo {year}
  {2020})}\BibitemShut {NoStop}%
\bibitem [{\citenamefont {{Romero-Shaw}~{\sl et al.}}(2020)}]{Bilby_ref}%
  \BibitemOpen
  \bibfield  {author} {\bibinfo {author} {\bibfnamefont {I.~M.}\ \bibnamefont
  {{Romero-Shaw}~{\sl et al.}}},\ }\href@noop {} {\bibfield  {journal}
  {\bibinfo  {journal} {Mon. Not. Roy. Astron. Soc.}\ }\textbf {\bibinfo
  {volume} {499}},\ \bibinfo {pages} {3295} (\bibinfo {year}
  {2020})}\BibitemShut {NoStop}%
\end{thebibliography}%
\end{document}